\documentclass[lettersize,journal]{IEEEtran}

\usepackage{amsmath,amsfonts,amssymb}
\usepackage{graphicx}
\usepackage[caption=false,font=normalsize,labelfont=sf,textfont=sf]{subfig}
\usepackage{textcomp}
\usepackage{stfloats}
\usepackage{url}
\usepackage{verbatim}
\usepackage{cite}
\usepackage{booktabs}
\usepackage{multirow}
\usepackage{rotating}
\usepackage{array}
\usepackage{makecell}
\usepackage{float}
\usepackage{amsthm}
\usepackage{algorithm}
\usepackage{algorithmic}

\usepackage{threeparttable}   

\begin{document}

\title{Two-Dimensional Graph Bi-Fractional Fourier Transform}

\author{Mingzhi~Wang and Zhichao~Zhang,~\IEEEmembership{Member,~IEEE}
\thanks{This work was supported in part by the Open Foundation of Hubei Key Laboratory of Applied Mathematics (Hubei University) under Grant HBAM202404; and in part by the Foundation of Key Laboratory of System Control and Information Processing, Ministry of Education under Grant Scip20240121. \emph{(Corresponding author: Zhichao~Zhang.)}}
\thanks{Mingzhi~Wang is with the School of Mathematics and Statistics, Nanjing University of Information Science and Technology, Nanjing 210044, China (e-mail: wmz200208@163.com).}
\thanks{Zhichao~Zhang is with the School of Mathematics and Statistics, Nanjing University of Information Science and Technology, Nanjing 210044, China, with the Hubei Key Laboratory of Applied Mathematics, Hubei University, Wuhan 430062, China, and also with the Key Laboratory of System Control and Information Processing, Ministry of Education, Shanghai Jiao Tong University, Shanghai 200240, China (e-mail: zzc910731@163.com).}}


\IEEEpubid{}

\maketitle

\begin{abstract}
	Graph signal processing (GSP) advances spectral analysis on irregular domains. However, existing two-dimensional graph fractional Fourier transform (2D-GFRFT) employs a single fractional order for both factor graphs, thereby limiting its adaptability to heterogeneous signals. We proposed the two-dimensional graph bi-fractional Fourier transform (2D-GBFRFT), which assigns independent fractional orders to the factor graphs of a Cartesian product while preserving separability. We established invertibility, unitarity, and index additivity, and developed two filtering schemes: a Wiener-style design through grid search and a differentiable framework that jointly optimizes transform orders and diagonal spectral filters. We further introduced a hybrid interpolation with the joint time-vertex fractional Fourier transform (JFRFT), controlled by a tunable parameter that balances the two methods. In the domains of synthetic Cartesian product graph signals, authentic temporal graph datasets, and dynamic image deblurring, 2D-GBFRFT consistently surpasses 2D-GFRFT and enhances JFRFT. Experimental results confirmed the versatility and superior performance of 2D-GBFRFT for filtering in GSP.
\end{abstract}


\bstctlcite{BSTcontrol}

\section{Introduction}
\begin{figure*}[!t]
  \centering
  \includegraphics[width=\textwidth]{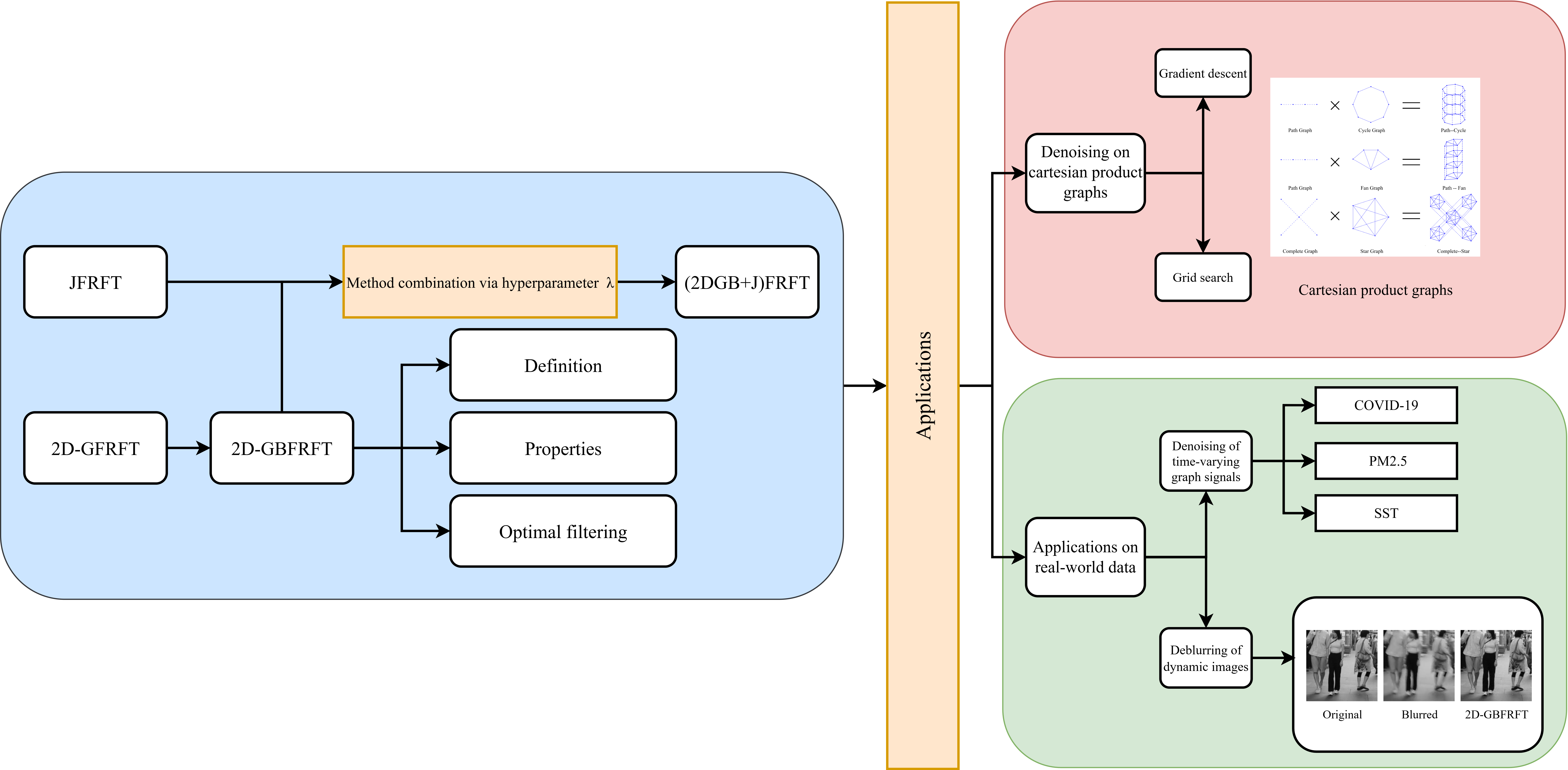} 
  \caption{Overview of the proposed 2D-GBFRFT framework and its applications.}
  \label{fig:flow3}
\end{figure*}
Signals originating from real-world systems are often characterized by complex networks or graphs, displaying non-Euclidean topologies. Examples encompass spatio-temporal data from environmental sensing networks~\cite{1,2,3}, interactions within social networks~\cite{4,5}, brain connectivity graphs in neuroscience~\cite{6,7}, and relationships among pixels or regions in image analysis~\cite{8,9}. These signals exist on irregular domains, complicating the direct application of classical techniques in both the time and frequency domains. Graph signal processing (GSP) offers a structured framework that extends traditional methodologies, including Fourier analysis~\cite{10,11,12,13,14}, wavelet transforms~\cite{15,16,17,18}, sampling theory~\cite{19,20,21}, and filtering~\cite{10},~\cite{12},~\cite{14},~\cite{22,23,24}, to graph domains, thereby facilitating efficient modeling and analysis on irregular topologies.

In GSP, the graph Fourier transform (GFT) represents signals in a spectral domain derived from the eigendecomposition of a graph matrix. Two common spectral bases are the graph Laplacian and the adjacency matrix. The Laplacian emphasizes smoothness and is suitable for diffusion-like processes, while the adjacency matrix captures connectivity, making it more effective for modeling nonstationary or locally high-frequency patterns~\cite{10},~\cite{11},~\cite{14},~\cite{25}. In this study, we use the adjacency matrix as the spectral basis to better model structurally complex signals with strong local variations. While GFT is effective, it relies on fixed integer-order spectral modes, which limits its ability to capture multiscale or heterogeneous structures.

To introduce spectral flexibility, the graph fractional Fourier transform (GFRFT) enhances GFT by incorporating a continuous spectral rotation parameter, facilitating smooth transformations between the vertex and graph-frequency domains. This extension has demonstrated efficacy in denoising, spectrum sparsification, and graph filtering~\cite{26,27,28,29,30,31,32}. Nevertheless, most existing GFRFT methods focus on single-graph signals and offer inadequate support for multidimensional or multi-structure graph signals.

In numerous applications, graph signals are intrinsically multidimensional, exemplified by time-vertex or space-attribute configurations. Cartesian product graphs offer an effective framework for modeling joint structures by integrating two factor graphs to represent multi-way dependencies. Utilizing this framework, the two-dimensional GFT (2D-GFT) executes distinct spectral decompositions of the factor graphs and integrates a unified spectrum through the Kronecker product, resulting in a separable frequency representation with advantageous computational characteristics~\cite{33,34,35}. Nonetheless, 2D-GFT retains the limitation of integer-order and encounters difficulties with heterogeneous spectral characteristics. The two-dimensional GFRFT (2D-GFRFT) resolves this issue by implementing a common fractional angle on both factor graphs, thereby facilitating continuous spectral control while maintaining computational efficiency~\cite{36}. This approach implicitly assumes that the spectra are similar across the factors. In situations characterized by significant heterogeneity, where the temporal dimension is uniform and the spatial dimension displays high-frequency fluctuations, its expressiveness may diminish.

Joint spectral analysis is essential for time-varying graph signals. The joint Fourier transform (JFT) integrates the discrete Fourier transform (DFT) in temporal domains with the GFT, facilitating the simultaneous modeling of temporal and structural dynamics\cite{37,38}. The joint time-vertex fractional Fourier transform (JFRFT) expands on this concept by introducing independent fractional angles for time and graph, enabling decoupled spectral rotations and increased flexibility~\cite{39,40,41}. JFRFT has demonstrated benefits in modeling spatio-temporal heterogeneity and improving denoising efficacy, prompting the advancement of more versatile two-dimensional spectral transforms.

Wiener filtering, based on the mean-squared error (MSE) criterion, is a commonly used denoising technique in spectral-domain filtering~\cite{14},~\cite{17},~\cite{42,43,44}. Its graphical counterpart employs spectral attenuation in a GFT basis and has been utilized for noise suppression in structured signals. GFRFT was subsequently utilized as a tunable spectral domain to more effectively align signal and noise spectra. Initial studies in GFRFT predominantly employed fixed fractional angles and derived filter coefficients from existing statistical knowledge~\cite{42}. For time-varying signals, JFRFT-based filters utilized two fractional angles and implemented grid search to determine the optimal parameters~\cite{39,40}. This method produced robust results; however, it relies on discrete parameter sweeps, which become computationally intensive as the dimensionality increases. Recent studies have presented differentiable formulations of GFRFT~\cite{29}, facilitating gradient-based training that optimizes both fractional angles and filter parameters concurrently within a cohesive differentiable architecture. Motivated by these advancements, learnable JFRFT-based filtering has been suggested for concurrent regulation in the time-vertex spectrum~\cite{41}. The results underscore the promise of fractional-domain designs for denoising and spectral shaping on graphs, illustrating how differentiable spectral transforms considerably enhance the expressiveness of traditional spectral methods.

The previous discussion elucidates both complementary strengths and limitations. The 2D-GFRFT is structurally uncomplicated and straightforward to implement; however, it depends on a singular fractional order, which may constrain its applicability for heterogeneous two-dimensional graph signals. JFRFT offers fractional angles for each dimension, enhancing flexibility in time-vertex analysis; nonetheless, it is constrained by particular time-space structures and does not directly extend to arbitrary Cartesian product graphs. These observations compel us to create a more cohesive fractional transform for general Cartesian product graphs. This novel transformation would concurrently maintain separability and computational efficiency, offer independent fractional tunability for each factor graph, and fulfill essential properties including invertibility, energy preservation, and exponential additivity. This transformation would provide adaptive spectral instruments for filtering and denoising on Cartesian product graphs, while also establishing the theoretical groundwork for multidimensional modeling in graph signal processing.

To address these challenges, we propose the two-dimensional graph bi-fractional Fourier transform (2D-GBFRFT). This approach incorporates fractional angles on the two-factor graphs of a Cartesian product graph, facilitating dimension-wise spectral rotations and heterogeneous control, while ensuring computational viability. In contrast to 2D-GFRFT, the 2D-GBFRFT alleviates the single-order assumption, facilitating a more precise representation of spectral characteristics per dimension. This is especially advantageous in heterogeneous situations where distinct dimensions display differing frequency characteristics. Unlike JFRFT, which is restricted to particular time-space frameworks, the 2D-GBFRFT is not bound by such limitations, rendering it suitable for a wider array of general Cartesian product graph signals.

\begin{itemize}
	\item We define the 2D-GBFRFT as a new spectral transform for Cartesian product graphs.
	\item We analyze the theoretical properties of the 2D-GBFRFT, including invertibility, degeneracy, and exponential additivity.
	\item We demonstrate the effectiveness of the 2D-GBFRFT in denoising multidimensional and heterogeneous graph signals.
	\item We introduce a parameterized interpolation between 2D-GBFRFT and JFRFT to improve flexibility in practical scenarios.
	\item We apply the proposed method to dynamic image deblurring, expanding its application potential in graph-based vision tasks.
\end{itemize}

The remainder of the study is organized as follows. Section~\ref{sec:prelim} reviews the preliminaries, including Cartesian product graphs, JFRFT, and 2D-GFRFT. Section~\ref{sec:2dgbfrft} introduces the 2D-GBFRFT, including its definition and theoretical properties. Section~\ref{sec:synthetic} examines the denoising performance of the 2D-GBFRFT on synthetic signals defined over Cartesian product graphs. Section~\ref{sec:realdata} extends the denoising method to real-world time-varying signals, proposes a hyperparameter \( \lambda \)-augmented extension to adaptively interpolate between the 2D-GBFRFT and JFRFT, referred to as (2DGB+J)FRFT, and demonstrates the practical effectiveness of the 2D-GBFRFT method on three real datasets, including its application to dynamic image deblurring. Finally, Section~\ref{sec:conclusion} concludes the study. Fig.~\ref{fig:flow3} presents an overview of the proposed 2D-GBFRFT framework and its applications. All the technical proofs of our theoretical results are relegated to the Appendix parts.

Notation: Bold lowercase letters denote vectors, and bold uppercase letters denote matrices. For a set $\mathcal{V}$, its cardinality is denoted by $|\mathcal{V}|$. The operators $(\cdot)^*$, $(\cdot)^T$, and $(\cdot)^H$ represent the complex conjugate, the transpose, and the Hermitian transpose, respectively. The identity matrix of size $N$ is denoted by $\mathbf{I}_N$. The operator $\mathrm{diag}(\mathbf{h})$ denotes the diagonal matrix whose diagonal entries are given by vector $\mathbf{h}$, while $\mathrm{diag}(\mathbf{H})$ denotes the vector of diagonal entries of matrix $\mathbf{H}$. The Frobenius norm of a matrix $\mathbf{A}$ is written as $\|\mathbf{A}\|_F$. The operator $\mathrm{vec}(\mathbf{A})$ denotes vectorization, obtained by stacking the columns of $\mathbf{A}$ into a long vector. The trace of a square matrix $\mathbf{A}$, denoted by $\operatorname{tr}(\mathbf{A})$, is the sum of its diagonal entries, i.e., $\operatorname{tr}(\mathbf{A}) = \sum_{i=1}^{n} A_{ii}$. For $\mathbf{A} \in \mathbb{C}^{m \times n}$ and $\mathbf{B} \in \mathbb{C}^{p \times q}$, the Kronecker product is denoted by $\mathbf{A} \otimes \mathbf{B} \in \mathbb{C}^{mp \times nq}$. For square matrices $\mathbf{A} \in \mathbb{C}^{n \times n}$ and $\mathbf{B} \in \mathbb{C}^{m \times m}$, the Kronecker sum is defined as $\mathbf{A} \oplus \mathbf{B} = \mathbf{A} \otimes \mathbf{I}_m + \mathbf{I}_n \otimes \mathbf{B}$. For two matrices of the same size, the Hadamard product is denoted by $\mathbf{A} \circ \mathbf{B}$.

\section{PRELIMINARIES}
\label{sec:prelim}
\subsection{Graph Signals}
Let \( G = (\mathcal{V}, \mathbf{A}) \) be a weighted undirected graph, where \( \mathcal{V} = \{v_1, v_2, \dots, v_N\} \) denotes the node set with cardinality \( |\mathcal{V}| = N \), and \( \mathbf{A} \in \mathbb{R}^{N \times N} \) is the weighted adjacency matrix. An entry \( \mathbf{A}_{ij} > 0 \) indicates that an edge exists between nodes \( v_i \) and \( v_j \) with weight \( \mathbf{A}_{ij} \), while \( \mathbf{A}_{ij} = 0 \) indicates the absence of an edge. A graph signal is a function \( x: \mathcal{V} \to \mathbb{R} \) that assigns a scalar value \( x_i \) to each node \( v_i \in \mathcal{V} \). The values of the signal are collected into the vector \( \mathbf{x} = [x_1, x_2, \dots, x_N]^T \in \mathbb{R}^N \)~\cite{45}, which represents the signal in the vertex domain, with structure governed by the connectivity encoded in \( \mathbf{A} \).

To incorporate temporal dynamics, we assume that the graph signal is observed at \( T \) successive time instants with a unit sampling interval. Let \( \mathbf{x}_t \in \mathbb{R}^N \) denote the signal at time \( t \). The time-varying graph signal is represented by the matrix \( \mathbf{X} = [\mathbf{x}_1, \mathbf{x}_2, \dots, \mathbf{x}_T] \in \mathbb{R}^{N \times T} \). Both \( \mathbf{X} \) and its vectorized form \( \mathbf{x} = \mathrm{vec}(\mathbf{X}) \in \mathbb{R}^{NT} \) are referred to as the time-vertex signal~\cite{46,47}.

A Cartesian product \( G_1 \square G_2 \) of two weighted graphs \( G_1 = (\mathcal{V}_1, \mathcal{E}_1, w_1) \) and \( G_2 = (\mathcal{V}_2, \mathcal{E}_2, w_2) \) is a graph with vertex set \( \mathcal{V}_1 \times \mathcal{V}_2 \). Its edge set \( \mathcal{E} \) is defined by
\begin{align*}
	\{(i_1, i_2), (j_1, j_2)\} \in \mathcal{E} \iff & \left[\{i_1, j_1\} \in \mathcal{E}_1,\, i_2 = j_2\right] \\
	& \text{or}\ [i_1 = j_1,\, \{i_2, j_2\} \in \mathcal{E}_2],
\end{align*}
and the weight function is given by
\[
w\big((i_1, i_2), (j_1, j_2)\big) 
= w_1(i_1, j_1)\,\delta(i_2, j_2) + \delta(i_1, j_1)\,w_2(i_2, j_2).
\]
Graphs \( G_1 \) and \( G_2 \) are referred to as the factor graphs of \( G_1 \square G_2 \).

Let \( G_n \) with vertex set \( \mathcal{V}_n = \{1,2,\dots,N_n\} \) have adjacency, degree, and Laplacian matrices \( \mathbf{W}_n \), \( \mathbf{D}_n \), and \( \mathbf{L}_n \), respectively, for \( n = 1, 2 \). When the vertices of the Cartesian product graph \( \mathcal{V}_1 \times \mathcal{V}_2 \) are ordered lexicographically, i.e., \( (1,1), (1,2), \dots, (N_1,N_2) \), the adjacency, degree, and Laplacian matrices of \( G_1 \square G_2 \) can be written as \( \mathbf{W}_1 \oplus \mathbf{W}_2 \), \( \mathbf{D}_1 \oplus \mathbf{D}_2 \), and \( \mathbf{L}_1 \oplus \mathbf{L}_2 \), respectively~\cite{33}.

\subsection{2D-GFRFT}

We consider the adjacency matrix \( \mathbf{A}_n \) of a weighted graph \( G_n \). 
For real symmetric weighted graphs, \( \mathbf{A}_n \) is diagonalizable with eigenvalue decomposition 
\( \mathbf{\Lambda}_{A_n} = \mathbf{V}_{A_n}^{\mathrm{H}} \mathbf{A}_n \mathbf{V}_{A_n} \), 
where \( \mathbf{V}_{A_n} \) contains the eigenvectors of \( \mathbf{A}_n \) and \( \mathbf{\Lambda}_{A_n} \) is a diagonal matrix of eigenvalues.
The GFT matrix is defined as \( \mathbf{F}_{G_n} = \mathbf{V}_{A_n}^{\mathrm{H}} \), 
and the inverse is given by \( \mathbf{F}_{G_n}^{-1} = \mathbf{V}_{A_n} \)~\cite{11}. 

We define the GFRFT as
\begin{equation}
	\mathbf{F}_{G_n}^{\alpha_n} = \mathbf{V}_{F_n} \, \mathbf{J}_{F_n}^{\alpha_n} \, \mathbf{V}_{F_n}^{-1},
	\label{eq:gfrft_def}
\end{equation}
where \( \mathbf{J}_{F_n}^{\alpha_n} \) represents the fractional matrix in the graph Fourier domain, and \( \mathbf{V}_{F_n} \) is the eigenvector matrix for \( G_n \).If \( \mathbf{F}_{G_n} \) is unitary, then \( (\mathbf{F}_{G_n}^{\alpha_n})^{\mathrm{H}} = \mathbf{F}_{G_n}^{-\alpha_n} \).

Given a graph signal $X: V_1 \times V_2 \to \mathbb{R}$ on the Cartesian product graph $G_1 \square G_2$,
let $N_i \triangleq |V_i|$ ($i=1,2$) and let $\mathbf{X}\in\mathbb{R}^{N_1\times N_2}$ denote the vertex-domain matrix representation of $X$.

\textit{Definition 1:}
The 2D-GFRFT of $\mathbf{X}$ is defined as
\begin{equation}
	\mathbf{X}_f^{\alpha}
	= \mathbf{F}_{G_1}^{\alpha}\,\mathbf{X}\,\big(\mathbf{F}_{G_2}^{\alpha}\big)^{\mathrm T},
	\label{eq:2dgfrft_matrix}
\end{equation}
where $\mathbf{F}_{G_i}^{\alpha}$ ($i=1,2$) are the $\alpha$-order GFRFT matrices of $G_i$.
The inverse transform is
\begin{equation}
	\mathbf{X}
	= \mathbf{F}_{G_1}^{-\alpha}\,\mathbf{X}_f^{\alpha}\,\big(\mathbf{F}_{G_2}^{-\alpha}\big)^{\mathrm T}.
	\label{eq:2dgfrft_matrix_inv}
\end{equation}

Equivalently, in vectorized form, letting
\( \mathbf{x} \triangleq \mathrm{vec}(\mathbf{X}) \) and
\( \mathbf{x}_f^{\alpha} \triangleq \mathrm{vec}(\mathbf{X}_f^{\alpha}) \), we have
\begin{equation}
	\mathbf{x}_f^{\alpha} = \big(\mathbf{F}_{G_2}^{\alpha} \otimes \mathbf{F}_{G_1}^{\alpha}\big)\,\mathbf{x}
	\;\triangleq\; \mathbf{F}_{2D}^{\alpha}\,\mathbf{x},
	\label{eq:2dgfrft_vec}
\end{equation}
with the corresponding inverse
\begin{equation}
	\mathbf{x} = \big(\mathbf{F}_{G_2}^{-\alpha} \otimes \mathbf{F}_{G_1}^{-\alpha}\big)\,\mathbf{x}_f^{\alpha}
	\;\triangleq\; \mathbf{F}_{2D}^{-\alpha}\,\mathbf{x}_f^{\alpha}.
	\label{eq:2dgfrft_vec_inv}
\end{equation}

The 2D-GFRFT thus provides a joint spectral representation spanned by the fractional eigenmodes of the two factor graphs and reduces to the classical 2D-GFT when \( \alpha = 1 \)~\cite{36}.

\subsection{JFRFT}

The fractional Fourier transform (FRFT) extends the classical Fourier transform by introducing a continuous order parameter \( \alpha \)~\cite{48,49,50}. Its discrete version, known as the discrete FRFT (DFRFT), is represented by a matrix \( \mathbf{F}^{\alpha} \in \mathbb{C}^{T \times T} \), which reduces to the DFT when \( \alpha = 1 \)~\cite{51}.

JFRFT generalizes this concept to time-varying graph signals, incorporating both the graph structure and temporal dynamics. Let \( \mathbf{X} \in \mathbb{C}^{N \times T} \) denote a signal matrix defined over a graph with \( N \) vertices and \( T \) time instants, and let \( \mathbf{x} = \mathrm{vec}(\mathbf{X}) \in \mathbb{C}^{NT} \) be its vectorized form.

\textit{Definition 2:} 
The JFRFT of $\mathbf{X}$ is defined as
\begin{equation}
	\mathbf{X}_{J}^{(\alpha,\beta)}
	= \mathbf{F}_{\mathrm G}^{\beta}\, \mathbf{X}\, \big(\mathbf{F}^{\alpha}\big)^{\mathrm T},
	\label{eq:jfrft_matrix}
\end{equation}
where $\mathbf{F}^{\alpha}$ is the $\alpha$-order DFRFT matrix, 
$\mathbf{F}_{\mathrm G}^{\beta}$ is the $\beta$-order GFRFT matrix, 
and $\mathbf{X}_{J}^{(\alpha,\beta)}$ denotes the JFRFT-domain representation of $\mathbf{X}$. 
The inverse transform is
\begin{equation}
	\mathbf{X}
	= \mathbf{F}_{\mathrm G}^{-\beta}\, \mathbf{X}_{J}^{(\alpha,\beta)}\, \big(\mathbf{F}^{-\alpha}\big)^{\mathrm T}.
	\label{eq:jfrft_matrix_inv}
\end{equation}

Equivalently, in vectorized form, we have
\begin{equation}
	\mathbf{x}_{J}^{(\alpha,\beta)}
	= \big(\mathbf{F}^{\alpha} \otimes \mathbf{F}_{\mathrm G}^{\beta}\big)\,\mathbf{x}
	\;\triangleq\; \mathbf{F}_{J}^{(\alpha,\beta)}\,\mathbf{x},
	\label{eq:jfrft_vec}
\end{equation}
with the corresponding inverse
\begin{equation}
	\mathbf{x}
	= \big(\mathbf{F}^{-\alpha} \otimes \mathbf{F}_{\mathrm G}^{-\beta}\big)\,\mathbf{x}_{J}^{(\alpha,\beta)}
	\;\triangleq\; \mathbf{F}_{J}^{(-\alpha,-\beta)}\,\mathbf{x}_{J}^{(\alpha,\beta)}.
	\label{eq:jfrft_vec_inv}
\end{equation}

The JFRFT thus provides a joint fractional spectral representation of time-varying graph signals and reduces to the classical JFT when $\alpha=1$ and $\beta=1$.

\section{Two-Dimensional Graph Bi-Fractional Fourier Transform (2D-GBFRFT)}
\label{sec:2dgbfrft}
\subsection{Definition}

The conventional 2D-GFRFT applies an identical fractional order to both factor graphs of a Cartesian product graph~\cite{36}. Although this uniform specification simplifies the transform, it may fail to capture structural heterogeneity between the two factors. In particular, when the factors exhibit significantly different characteristics, such as in space-time graph scenarios, a single shared order can introduce spectral aliasing or obscure directional information.

To overcome these limitations, we propose the 2D-GBFRFT. Unlike the conventional formulation, 2D-GBFRFT assigns two independent fractional orders, \( \alpha_1 \) and \( \alpha_2 \), to the factor graphs, enabling adaptivity along each dimension. This bi-fractional design improves both the flexibility and fidelity of spectral representations, particularly when the factor graphs have distinct structural properties.

Let two graphs be given as $G_1 = (\mathcal{V}_1, \mathcal{E}_1, \mathbf{W}_1)$ and $G_2 = (\mathcal{V}_2, \mathcal{E}_2, \mathbf{W}_2)$, with $N_1=|\mathcal{V}_1|$ and $N_2=|\mathcal{V}_2|$. Their adjacency matrices $\mathbf{A}_1$ and $\mathbf{A}_2$ admit the eigendecompositions
\begin{equation}
	\mathbf{A}_1 = \boldsymbol{\chi}_1 \boldsymbol{\Lambda}_1 \boldsymbol{\chi}_1^{\mathrm{T}},
	\quad
	\mathbf{A}_2 = \boldsymbol{\chi}_2 \boldsymbol{\Lambda}_2 \boldsymbol{\chi}_2^{\mathrm{T}}.
	\label{eq:2dgbfrft_lap}
\end{equation}
Given two independent fractional orders $\alpha_1, \alpha_2 \in \mathbb{R}$, the bi-fractional transform matrices are defined as
\begin{equation}
	\mathbf{F}_{G_1}^{\alpha_1} = \boldsymbol{\chi}_1 \boldsymbol{\Lambda}_1^{\alpha_1} \boldsymbol{\chi}_1^{\mathrm{T}},
	\quad
	\mathbf{F}_{G_2}^{\alpha_2} = \boldsymbol{\chi}_2 \boldsymbol{\Lambda}_2^{\alpha_2} \boldsymbol{\chi}_2^{\mathrm{T}}.
	\label{eq:2dgbfrft_fg}
\end{equation}

Given a graph signal $X: \mathcal{V}_1 \times \mathcal{V}_2 \to \mathbb{R}$ on the Cartesian product graph $G_1 \square G_2$, let $N_i \triangleq |\mathcal{V}_i|$ ($i=1,2$) and let $\mathbf{X}\in\mathbb{R}^{N_1\times N_2}$ denote the vertex-domain matrix representation of $X$.

\textit{Definition 3:}
The 2D-GBFRFT of $\mathbf{X}$ is defined as
\begin{equation}
	\mathbf{X}_f^{\alpha_1,\alpha_2}
	= \mathbf{F}_{G_1}^{\alpha_1}\,\mathbf{X}\,\big(\mathbf{F}_{G_2}^{\alpha_2}\big)^{\mathrm T},
	\label{eq:2dgbfrft_matrix}
\end{equation}
where $\mathbf{F}_{G_i}^{\alpha_i}$ ($i=1,2$) denote the $\alpha_i$-order GFRFT matrices of $G_i$, and $\mathbf{X}_f^{\alpha_1,\alpha_2}$ is the 2D-GBFRFT representation of $\mathbf{X}$.
The inverse transform is
\begin{equation}
	\mathbf{X}
	= \mathbf{F}_{G_1}^{-\alpha_1}\,\mathbf{X}_f^{\alpha_1,\alpha_2}\,\big(\mathbf{F}_{G_2}^{-\alpha_2}\big)^{\mathrm T}.
	\label{eq:2dgbfrft_matrix_inv}
\end{equation}

In vectorized form, letting $\mathbf{x}\triangleq\mathrm{vec}(\mathbf{X})$ and $\mathbf{x}_f^{\alpha_1,\alpha_2}\triangleq\mathrm{vec}(\mathbf{X}_f^{\alpha_1,\alpha_2})$, we have
\begin{equation}
	\mathbf{x}_f^{\alpha_1,\alpha_2}
	= \big(\mathbf{F}_{G_2}^{\alpha_2}\otimes \mathbf{F}_{G_1}^{\alpha_1}\big)\,\mathbf{x}
	\;\triangleq\; \mathbf{F}_{2D}^{(\alpha_1,\alpha_2)}\,\mathbf{x},
	\label{eq:2dgbfrft_vec}
\end{equation}
with the corresponding inverse
\begin{equation}
	\mathbf{x}
	= \big(\mathbf{F}_{G_2}^{-\alpha_2}\otimes \mathbf{F}_{G_1}^{-\alpha_1}\big)\,\mathbf{x}_f^{\alpha_1,\alpha_2}
	\;\triangleq\; \mathbf{F}_{2D}^{(-\alpha_1,-\alpha_2)}\,\mathbf{x}_f^{\alpha_1,\alpha_2}.
	\label{eq:2dgbfrft_vec_inv}
\end{equation}

\subsection{Properties}

The proposed 2D-GBFRFT inherits and extends several classical properties of FRFT and GFRFT. 
We summarize the main theoretical properties that establish its effectiveness in graph signal processing.

\textit{Property 1 (Identity):}  
If $\alpha_1 = \alpha_2 = 0$, the 2D-GBFRFT reduces to the identity transform, i.e.,
\begin{equation}
\mathbf{F}_{2D}^{(0,0)} 
= \mathbf{F}_{G_2}^{0} \otimes \mathbf{F}_{G_1}^{0} 
= \mathbf{I}_{N_2} \otimes \mathbf{I}_{N_1} 
= \mathbf{I}_{N_1N_2}.
\label{eq:2dgbfrft_identity}
\end{equation}

\textit{Property 2 (Reduction to 2D-GFRFT):}  
When $\alpha_1 = \alpha_2 = \alpha$, the 2D-GBFRFT reduces to the standard 2D-GFRFT:
\begin{equation}
\mathbf{F}_{2D}^{(\alpha,\alpha)} 
= \mathbf{F}_{G_2}^{\alpha} \otimes \mathbf{F}_{G_1}^{\alpha} 
= \mathbf{F}_{2D}^{\alpha}.
\label{eq:2dgbfrft_reduction}
\end{equation}

\textit{Property 3 (Unitarity):}  
If both $\mathbf{F}_{G_1}^{\alpha_1}$ and $\mathbf{F}_{G_2}^{\alpha_2}$ are unitary, 
then $\mathbf{F}_{2D}^{(\alpha_1,\alpha_2)}$ is also unitary. Specifically,
\begin{align}
	\big(\mathbf{F}_{2D}^{(\alpha_1,\alpha_2)}\big)^{\mathrm H}\,
	\mathbf{F}_{2D}^{(\alpha_1,\alpha_2)}
	&= \big(\mathbf{F}_{G_2}^{\alpha_2} \otimes \mathbf{F}_{G_1}^{\alpha_1}\big)^{\mathrm H}
	\big(\mathbf{F}_{G_2}^{\alpha_2} \otimes \mathbf{F}_{G_1}^{\alpha_1}\big) \notag\\
	&= \left( \bigl(\mathbf{F}_{G_2}^{\alpha_2}\bigr)^{\mathrm H}\mathbf{F}_{G_2}^{\alpha_2} \right)
	\ \otimes\
	\left( \bigl(\mathbf{F}_{G_1}^{\alpha_1}\bigr)^{\mathrm H}\mathbf{F}_{G_1}^{\alpha_1} \right) \notag\\
	&= \mathbf{I}_{N_1N_2},
	\label{eq:2dgbfrft_unitarity}
\end{align}

which establishes unitarity.

\textit{Property 4 (Index additivity):}  
For any two sets of orders $(\alpha_1,\alpha_2)$ and $(\beta_1,\beta_2)$, the additivity property holds:
\begin{align}
\mathbf{F}_{2D}^{(\alpha_1,\alpha_2)} \mathbf{F}_{2D}^{(\beta_1,\beta_2)}
&= (\mathbf{F}_{G_2}^{\alpha_2} \otimes \mathbf{F}_{G_1}^{\alpha_1})
   (\mathbf{F}_{G_2}^{\beta_2} \otimes \mathbf{F}_{G_1}^{\beta_1}) \notag \\
&= (\mathbf{F}_{G_2}^{\alpha_2}\mathbf{F}_{G_2}^{\beta_2})
   \otimes
   (\mathbf{F}_{G_1}^{\alpha_1}\mathbf{F}_{G_1}^{\beta_1}) \notag \\
&= \mathbf{F}_{G_2}^{\alpha_2+\beta_2} \otimes \mathbf{F}_{G_1}^{\alpha_1+\beta_1} \notag \\
&= \mathbf{F}_{2D}^{(\alpha_1+\beta_1,\,\alpha_2+\beta_2)}.
\label{eq:2dgbfrft_additivity}
\end{align}
In particular, setting $(\beta_1,\beta_2) = (-\alpha_1,-\alpha_2)$ yields
\begin{equation}
\mathbf{F}_{2D}^{(\alpha_1,\alpha_2)} 
\mathbf{F}_{2D}^{(-\alpha_1,-\alpha_2)}
= \mathbf{F}_{2D}^{(0,0)} 
= \mathbf{I}_{N_1N_2},
\label{eq:2dgbfrft_invertibility}
\end{equation}
which confirms the invertibility of the 2D-GBFRFT.

\begin{algorithm}[t]
	\caption{2D-GBFRFT Order Selection via Grid Search}
	\label{alg:gridsearch}
	\begin{algorithmic}[1]
		\REQUIRE 
		\( \mathbf{y} = (\mathbf{G}_{G_2}^\top \otimes \mathbf{G}_{G_1}) \mathbf{x} + \mathbf{n} \);
		Statistics \( \mathbb{E}[\mathbf{x} \mathbf{x}^{\mathrm{H}}] \), 
		\( \mathbb{E}[\mathbf{n} \mathbf{n}^{\mathrm{H}}] \),  
		Ranges \( \alpha_1 \in [a, b] \), \( \alpha_2 \in [c, d] \)
		\ENSURE Selected \( (\alpha_1^\ast,\alpha_2^\ast) \) and corresponding MSE \( \mathrm{MSE}^\ast \)
		
		\FOR{$\alpha_1 \in [a,b]$}
		\FOR{$\alpha_2 \in [c,d]$}
		\STATE \textbf{Step 1: Operator construction}
		\STATE Compute \( \mathbf{F}_{G_1}^{\alpha_1} \), \( \mathbf{F}_{G_2}^{\alpha_2} \); set
		\( \mathbf{F}_{2D}^{(\alpha_1,\alpha_2)} = \mathbf{F}_{G_2}^{\alpha_2} \otimes \mathbf{F}_{G_1}^{\alpha_1} \)
		\STATE \textbf{Step 2: Filter computation}
		\STATE Form \( \{ \mathbf{W}_m \}_{m=0}^{N_1N_2-1} \) from rows/columns of \( \mathbf{F}_{2D}^{(\alpha_1, \alpha_2)} \)
		\STATE Assemble \( \mathbf{T} \), \( \mathbf{q} \) using trace expressions
		\STATE Solve \( \mathbf{h}^{\mathrm{opt}} = \mathbf{T}^{-1} \mathbf{q} \)
		\STATE \textbf{Step 3: MSE evaluation and update}
		\STATE Compute MSE and update \( (\alpha_1^\ast, \alpha_2^\ast), \ \mathrm{MSE}^\ast \) if this value is smaller than the current best
		\ENDFOR
		\ENDFOR
		\STATE \textbf{Step 4: Store results}
		\STATE Return \( (\alpha_1^\ast, \alpha_2^\ast) \) and \( \mathrm{MSE}^\ast \)
	\end{algorithmic}
\end{algorithm}

\subsection{Optimal Filtering in 2D-GBFRFT Domains}

In JFRFT domains,the Wiener filtering problem was considered in~\cite{40} for time-vertex signals.We generalize the Wiener filtering problem to Cartesian product graph signals as follows.We consider a stochastic signal observation model over a Cartesian product graph \( G_1 \square G_2 \), described as
\begin{equation}
\mathbf{Y} = \mathbf{G}_{G_1} \mathbf{X} \mathbf{G}_{G_2} + \mathbf{N},
\label{eq:obs_model_matrix}
\end{equation}
where \( \mathbf{G}_{G_1} \in \mathbb{C}^{N_1 \times N_1} \) and \( \mathbf{G}_{G_2} \in \mathbb{C}^{N_2 \times N_2} \) are known graph filters corresponding to the factor graphs \( G_1 \) and \( G_2 \), respectively. The matrix \( \mathbf{X} \in \mathbb{C}^{N_1 \times N_2} \) represents a stochastic graph signal defined on \( G_1 \square G_2 \), and \( \mathbf{N} \in \mathbb{C}^{N_1 \times N_2} \) denotes an additive noise term. Denoting \( \mathbf{x} = \mathrm{vec}(\mathbf{X}) \) and \( \mathbf{n} = \mathrm{vec}(\mathbf{N}) \), the observation model can be rewritten in vectorized form as
\begin{equation}
\mathbf{y} = (\mathbf{G}_{G_2}^\top \otimes \mathbf{G}_{G_1}) \mathbf{x} + \mathbf{n},
\label{eq:obs_model_vector}
\end{equation}
where \( \mathbf{y} = \mathrm{vec}(\mathbf{Y}) \).

We assume the first- and second-order statistical properties of both signal and noise are known, including the expectations \( \mathbb{E}\{\mathbf{x}\}, \mathbb{E}\{\mathbf{n}\} \), covariances \( \mathbb{E}\{\mathbf{x} \mathbf{x}^{\mathrm{H}}\}, \mathbb{E}\{\mathbf{n} \mathbf{n}^{\mathrm{H}}\} \), and cross-covariances \( \mathbb{E}\{\mathbf{x} \mathbf{n}^{\mathrm{H}}\}, \mathbb{E}\{\mathbf{n} \mathbf{x}^{\mathrm{H}}\} \).

\subsubsection{Spectral Filtering under 2D-GBFRFT via Grid Search}

\hspace*{1em}\textit{Theorem 1:}  
 Given the observation model in \eqref{eq:obs_model_vector} and the 2D-GBFRFT operator \( \mathbf{F}_{2D}^{(\alpha_1,\alpha_2)} \), the optimal diagonal filter \( \mathbf{H}_{2D} \) that minimizes the MSE 
\begin{equation}
	\mathbb{E} \left\{ \left\| \mathbf{F}_{2D}^{(-\alpha_1, -\alpha_2)} \mathbf{H}_{2D} \mathbf{F}_{2D}^{(\alpha_1, \alpha_2)} \mathbf{y} - \mathbf{x} \right\|_2^2 \right\}
\end{equation}
is given by the solution to the linear system
\begin{equation}
\mathbf{T} \mathbf{h}^{\mathrm{opt}} = \mathbf{q}.
\end{equation}
The detailed expressions for the matrix \( \mathbf{T} \) and vector \( \mathbf{q} \) are provided in the proof below.

\textit{Proof:} See Appendix~A.\qed

The filtering performance depends on the choice of fractional orders $(\alpha_1,\alpha_2)$. 
To select the optimal pair, we perform a grid search over $\alpha_1\in[a, b]$ and $\alpha_2\in[c, d]$. 
For each candidate $(\alpha_1,\alpha_2)$, we form the 2D-GBFRFT operator $\mathbf{F}_{2D}^{(\alpha_1,\alpha_2)}$, solve for the optimal filter coefficients $\mathbf{h}^{\mathrm{opt}}$, and compute the resulting MSE. 
We then select the pair that minimizes the MSE.

We remark that the expressions in \eqref{eq:T_entries} and \eqref{eq:q_entries} contain cross-terms between signal and noise. Under the common assumption of independence, these cross-terms vanish, which simplifies the computation of \(\mathbf{T}\) and \(\mathbf{q}\) and reduces overall complexity. The complete procedure is presented as pseudocode in Algorithm~\ref{alg:gridsearch}.
\subsubsection{Spectral Filtering under 2D-GBFRFT via Gradient Descent}

Instead of fixing the transform orders $(\alpha_1,\alpha_2)$ by grid search, 
we jointly optimize the orders and the diagonal spectral filter 
by directly minimizing the MSE inspired by~\cite{29}.

\textit{Theorem 2:}  
The 2D-GBFRFT operator \( \mathbf{F}_{2D}^{(\alpha_1, \alpha_2)} \) is differentiable with respect to \( \alpha_1 \) and \( \alpha_2 \). This differentiability enables joint optimization via gradient descent. The learning objective is
\begin{equation}
	\begin{aligned}
		\min_{\alpha_1,\alpha_2,\,\mathbf h} 
		\mathcal L(\alpha_1,\alpha_2,\mathbf h) \\  
		&\hspace{-7em}= \mathbb{E}\!\left\{ \left\|  
		\mathbf{F}_{2D}^{(-\alpha_1,-\alpha_2)} \mathbf{H}_{2D} 
		\mathbf{F}_{2D}^{(\alpha_1,\alpha_2)} \mathbf y - \mathbf x \right\|_2^2 
		\right\}
	\end{aligned}
	\label{eq:obj-2dgb}
\end{equation}
where \( \mathbf{H}_{2D} = \mathrm{diag}(\mathbf h) \).

The update rules for the optimization are as follows:
\begin{equation}
	\begin{bmatrix}
		\alpha_1^{\text{next}}\\
		\alpha_2^{\text{next}}
	\end{bmatrix}
	= 
	\begin{bmatrix}
		\alpha_1^{\text{current}}\\
		\alpha_2^{\text{current}}
	\end{bmatrix}
	- \gamma
	\begin{bmatrix}
		\frac{\partial \mathcal{L}}{\partial \alpha_1} \\
		\frac{\partial \mathcal{L}}{\partial \alpha_2}
	\end{bmatrix},
	\qquad 
	\mathbf{h} \leftarrow \mathbf{h} - \gamma_h \frac{\partial \mathcal{L}}{\partial \mathbf{h}}.
	\label{eq:update-rules}
\end{equation}
Here, \( \gamma \) and \( \gamma_h \) are the learning rates for \( \alpha_1, \alpha_2 \) and \( \mathbf{h} \), respectively.

\textit{Proof:} See Appendix~B.\qed

Compared with grid search, this formulation treats $(\alpha_1,\alpha_2)$ as continuous trainable variables optimized jointly with $\mathbf h$, thereby avoiding exhaustive enumeration over the parameter grid. In practice, implementation requires initialization of $(\alpha_1,\alpha_2)$, distinct learning rates for the transform orders and the filter coefficients, and an appropriate stopping criterion to ensure stable convergence. The estimator is given by
\[
\hat{\mathbf{x}} = \mathbf{F}_{2D}^{(-\alpha_1, -\alpha_2)} \mathbf{H}_{2D} \mathbf{F}_{2D}^{(\alpha_1, \alpha_2)} \mathbf{y}.
\]

The complete procedure is summarized in Algorithm~\ref{alg:gd-2dgbfrft}.

\begin{algorithm}[htbp]
\caption{2D-GBFRFT Order and Filter Optimization via Gradient Descent}
\label{alg:gd-2dgbfrft}
\begin{algorithmic}[1]
\REQUIRE  $\mathbf{y}=(\mathbf{G}_{G_2}^{\top}\!\otimes\!\mathbf{G}_{G_1})\mathbf{x}+\mathbf{n}$, 
          learning rates $\gamma$, training epochs $L$
\ENSURE Trained orders $(\alpha_1,\alpha_2)$, diagonal filter $\mathbf{H}_{2D}$, final MSE

\STATE Initialize $\alpha_1=\alpha_2=0.5$, $\mathbf{H}_{2D}=\mathbf{I}$
\FOR{$t=1$ to $L$}
  \STATE \textbf{Step 1: Operator construction}
  \STATE Build $\mathbf{F}_{2D}^{(\alpha_1,\alpha_2)}=\mathbf{F}_{G_2}^{\alpha_2}\otimes\mathbf{F}_{G_1}^{\alpha_1}$
  \STATE \textbf{Step 2: Forward propagation}
  \STATE $\hat{\mathbf{x}}=\mathbf{F}_{2D}^{-1}\mathbf{H}_{2D}\mathbf{F}_{2D}\mathbf{y}$
  \STATE \textbf{Step 3: Loss calculation and update}
  \STATE Compute $\mathcal{L}=\|\hat{\mathbf{x}}-\mathbf{x}\|_2^2$
  \STATE Update $(\alpha_1,\alpha_2,\mathbf{H}_{2D})$ by gradient descent
\ENDFOR
\STATE \textbf{Step 4: Store results}
\STATE Return $(\alpha_1,\alpha_2)$, $\mathbf{H}_{2D}$, and final MSE
\end{algorithmic}
\end{algorithm}

\subsubsection{Computational Complexity Analysis}
In the grid search method for 2D-GBFRFT, GFRFT matrices \( \mathbf{F}_{G_1}^{\alpha_1} \) and \( \mathbf{F}_{G_2}^{\alpha_2} \) must be computed 
for every candidate pair \( (\alpha_1, \alpha_2) \) in the search range. 
This requires eigenvalue or Jordan-type decompositions of the graph shift operators, 
which have a computational complexity of \( O(N_1^3 + N_2^3) \), where \( N_1 \) and \( N_2 \) represent 
the numbers of vertices in graphs \( G_1 \) and \( G_2 \), respectively. 
Additionally, for each pair \( (\alpha_1, \alpha_2) \), constructing the basis matrices 
and solving for the optimal filter coefficients involves solving a large linear system, 
which has a complexity of \( O(N_1^4 N_2^4) \). 
Therefore, the overall complexity of the grid search approach is 
\( O(N_1^3 + N_2^3 + N_1^4 N_2^4) \).  

In contrast, the gradient descent method also requires an initial decomposition 
of the graph operators to construct the fractional bases. 
However, once the bases are computed, the transform orders \( (\alpha_1, \alpha_2) \) 
and diagonal filter \( \mathbf{H}_{2D} \) are updated dynamically during training. 
Each iteration involves matrix–vector multiplications and diagonal updates, 
with per-epoch complexity \( O(N_1^2 N_2^2) \). 
Thus, the total complexity of the gradient descent method is 
\( O(N_1^3 + N_2^3 + N_1^2 N_2^2) \), 
which offers significantly improved computational efficiency compared 
to the conventional grid search approach. The comparison of computational complexity between grid search and gradient descent is summarized in Table~\ref{tab:comparison}.
\begin{table}[t]
	\caption{Comparison of Computational Complexity between 2D-GBFRFT via Grid Search and Gradient Descent}
	\label{tab:comparison}
	\centering
	\begin{tabular}{lcc}
		\toprule
		Method & Parameter Count & Complexity \\
		\midrule
		2D-GBFRFT\textsubscript{GS} & \( \left(\frac{b-a}{\Delta \alpha_1} + 1\right)\left(\frac{d-c}{\Delta \alpha_2} + 1\right) \) & \( O(N_1^3 + N_2^3 + N_1^4 N_2^4) \) \\
		2D-GBFRFT\textsubscript{GD} & \( (N_1 N_2 + 2) n_{\exp} \) & \( O(N_1^3 + N_2^3 + N_1^2 N_2^2) \) \\
		\bottomrule
	\end{tabular}
	
	\vspace{6pt}
	\footnotesize
	\begin{tabular}{@{}p{\columnwidth}@{}}
		\textbf{Note:} GS denotes grid search and GD denotes gradient descent;
		\( N_1 \), \( N_2 \) are the numbers of vertices in \( G_1 \) and \( G_2 \), respectively;
		\( n_{\exp} \) is the number of experimental runs with different initializations;
		\( [a, b] \) and \( [c, d] \) are the search ranges of the transform orders \( \alpha_1 \) and \( \alpha_2 \);
		\( \Delta\alpha_1 \), \( \Delta\alpha_2 \) are the corresponding grid step sizes.
	\end{tabular}
\end{table}

\subsubsection{Convexity and Convergence Analysis}

Grid search and gradient descent exhibit fundamental differences in convexity and convergence guarantees. In grid search, the transform orders \( (\alpha_1, \alpha_2) \) are treated as hyperparameters. For each fixed pair, the filter coefficients \( \mathbf{h} \) are obtained in closed form by solving the Wiener-type linear system. This corresponds to minimizing a quadratic MSE criterion, which is convex in \( \mathbf{h} \), thus guaranteeing a global optimum. Therefore, grid search can be viewed as performing multiple independent convex optimizations, each associated with a candidate pair \( (\alpha_1, \alpha_2) \)~\cite{42}. Within the finite search range \( [a, b] \times [c, d] \), the solution corresponds to the global minimizer of the MSE over all candidates. The trade-off involves the grid resolution: coarse grids provide lower precision but lower computational cost, whereas finer grids improve precision but increase computational expense.

In contrast, gradient descent jointly optimizes \( (\alpha_1, \alpha_2) \) and \( \mathbf{h} \). While the subproblem in \( \mathbf{h} \) remains convex, the overall optimization is nonconvex due to the nonlinear dependence of the 2D-GBFRFT operator \( \mathbf{F}_{2D}^{(\alpha_1, \alpha_2)} \) on its fractional orders. As a result, the MSE surface may contain multiple local minima, and convergence is only guaranteed to a stationary point. The outcome strongly depends on initialization, step sizes, and problem conditioning. However, in practice, gradient descent often identifies high-quality solutions efficiently, especially when the MSE landscape has broad basins of attraction around good optima.

From a theoretical standpoint, grid search guarantees global optimality within a bounded discrete range, while gradient descent provides local optimality in the continuous domain. From a practical perspective, grid search is exhaustive and robust but computationally expensive, whereas gradient descent is efficient and adaptive, yet sensitive to initialization and hyperparameters. Overall, the two strategies highlight a trade-off between theoretical guarantees and computational feasibility: grid search is preferable in low-dimensional, well-bounded spaces, while gradient descent is more suitable for high-resolution or high-dimensional settings, where exhaustive enumeration is impractical.

\section{Signal Denoising on Cartesian Product Graphs}
\label{sec:synthetic}

\begin{table*}[htbp]
	\centering
	\caption{DENOISING RESULTS (MSE) ON CARTESIAN PRODUCT GRAPHS USING GRID SEARCH METHOD}
	\label{tab:grid}
	\resizebox{\textwidth}{!}{%
		\scriptsize
		\begin{tabular}{@{}c*{9}{c}@{}}
			\toprule
			\multirow{4}{*}{Method} & \multicolumn{3}{c}{Path-Cycle} & \multicolumn{3}{c}{Path-Fan} & \multicolumn{3}{c}{Complete-Star} \\
			\cmidrule(lr){2-4} \cmidrule(lr){5-7} \cmidrule(lr){8-10}
			& $\sigma=0.5$ & $\sigma=1.0$ & $\sigma=1.5$
			& $\sigma=0.5$ & $\sigma=1.0$ & $\sigma=1.5$
			& $\sigma=0.5$ & $\sigma=1.0$ & $\sigma=1.5$ \\
			\midrule
			2D-GFRFT\textsubscript{UU}
			& \makecell{$10.6529$ \\ $(0.7)$} & \makecell{$10.6999$ \\ $(0.7)$} & \makecell{$10.7659$ \\ $(0.7)$}
			& \makecell{$5.2421$ \\ $(0.8)$} & \makecell{$5.334$ \\ $(0.8)$} & \makecell{$5.4485$ \\ $(0.8)$}
			& \makecell{$5.5547$ \\ $(0.5)$} & \makecell{$5.6506$ \\ $(0.5)$} & \makecell{$5.7575$ \\ $(0.5)$} \\
			2D-GBFRFT\textsubscript{UU}
			& \makecell{$\mathbf{10.2136}$ \\ $(0.80, 0.30)$} & \makecell{$\mathbf{10.301}$ \\ $(0.80, 0.30)$} & \makecell{$\mathbf{10.4201}$ \\ $(0.80, 0.30)$}
			& \makecell{$\mathbf{5.1452}$ \\ $(0.90, 0.60)$} & \makecell{$\mathbf{5.2468}$ \\ $(0.90, 0.60)$} & \makecell{$\mathbf{5.3735}$ \\ $(0.90, 0.60)$}
			& \makecell{$\mathbf{5.4379}$ \\ $(0.70, 0.10)$} & \makecell{$\mathbf{5.5573}$ \\ $(0.70, 0.10)$} & \makecell{$\mathbf{5.6876}$ \\ $(0.70, 0.10)$} \\
			2D-GFRFT\textsubscript{UW}
			& \makecell{$10.6233$ \\ $(0.2)$} & \makecell{$10.6674$ \\ $(0.2)$} & \makecell{$10.7299$ \\ $(0.2)$}
			& \makecell{$4.9895$ \\ $(0.5)$} & \makecell{$5.1172$ \\ $(0.5)$} & \makecell{$5.2727$ \\ $(0.5)$}
			& \makecell{$5.578$ \\ $(0.2)$} & \makecell{$5.655$ \\ $(0.2)$} & \makecell{$5.7492$ \\ $(0.2)$} \\
			2D-GBFRFT\textsubscript{UW}
			& \makecell{$\mathbf{10.6133}$ \\ $(0.20, 0.30)$} & \makecell{$\mathbf{10.6598}$ \\ $(0.20, 0.30)$} & \makecell{$\mathbf{10.7253}$ \\ $(0.20, 0.30)$}
			& \makecell{$\mathbf{4.9418}$ \\ $(0.50, 0.60)$} & \makecell{$\mathbf{5.0752}$ \\ $(0.50, 0.60)$} & \makecell{$\mathbf{5.2369}$ \\ $(0.50, 0.60)$}
			& \makecell{$\mathbf{5.5337}$ \\ $(0.20, 0.40)$} & \makecell{$\mathbf{5.612}$ \\ $(0.20, 0.40)$} & \makecell{$\mathbf{5.7088}$ \\ $(0.20, 0.40)$} \\
			2D-GFRFT\textsubscript{DU}
			& \makecell{$30.7161$ \\ $(0.3)$} & \makecell{$30.7479$ \\ $(0.3)$} & \makecell{$30.7975$ \\ $(0.3)$}
			& \makecell{$18.735$ \\ $(0.5)$} & \makecell{$18.7735$ \\ $(0.5)$} & \makecell{$18.8321$ \\ $(0.5)$}
			& \makecell{$8.2204$ \\ $(0.7)$} & \makecell{$8.2737$ \\ $(0.7)$} & \makecell{$8.3471$ \\ $(0.7)$} \\
			2D-GBFRFT\textsubscript{DU}
			& \makecell{$\mathbf{19.5135}$ \\ $(0.70, 0.30)$} & \makecell{$\mathbf{19.5848}$ \\ $(0.60, 0.30)$} & \makecell{$\mathbf{19.6907}$ \\ $(0.60, 0.30)$}
			& \makecell{$\mathbf{7.864}$ \\ $(0.20, 0.90)$} & \makecell{$\mathbf{7.9397}$ \\ $(0.20, 0.90)$} & \makecell{$\mathbf{8.0365}$ \\ $(0.20, 0.90)$}
			& \makecell{$\mathbf{8.2162}$ \\ $(0.80, 0.70)$} & \makecell{$\mathbf{8.2694}$ \\ $(0.80, 0.70)$} & \makecell{$\mathbf{8.3427}$ \\ $(0.80, 0.70)$} \\
			2D-GFRFT\textsubscript{DW}
			& \makecell{$30.7548$ \\ $(0.7)$} & \makecell{$30.7821$ \\ $(0.7)$} & \makecell{$30.8248$ \\ $(0.7)$}
			& \makecell{$19.0336$ \\ $(0.4)$} & \makecell{$19.0787$ \\ $(0.4)$} & \makecell{$19.1411$ \\ $(0.3)$}
			& \makecell{$7.9615$ \\ $(0.4)$} & \makecell{$8.0382$ \\ $(0.3)$} & \makecell{$8.1408$ \\ $(0.3)$} \\
			2D-GBFRFT\textsubscript{DW}
			& \makecell{$\mathbf{19.9187}$ \\ $(0.60, 0.70)$} & \makecell{$\mathbf{19.9718}$ \\ $(0.60, 0.70)$} & \makecell{$\mathbf{20.0512}$ \\ $(0.60, 0.70)$}
			& \makecell{$\mathbf{8.2119}$ \\ $(0.40, 0.10)$} & \makecell{$\mathbf{8.2908}$ \\ $(0.30, 0.10)$} & \makecell{$\mathbf{8.3971}$ \\ $(0.30, 0.10)$}
			& \makecell{$\mathbf{7.7571}$ \\ $(0.30, 0.70)$} & \makecell{$\mathbf{7.8546}$ \\ $(0.30, 0.70)$} & \makecell{$\mathbf{7.9845}$ \\ $(0.30, 0.70)$} \\
			\bottomrule
		\end{tabular}
	}
	\\[0.5ex]
	\footnotesize \textbf{Note:} The subscripts indicate graph types: 
	UU = undirected unweighted, 
	UW = undirected weighted, 
	DU = directed unweighted, 
	DW = directed weighted.
\end{table*}

\begin{table*}[htbp]
	\centering
	\caption{DENOISING RESULTS (MSE) ON CARTESIAN PRODUCT GRAPHS USING GRADIENT DESCENT}
	\label{tab:grad}
	\setlength{\heavyrulewidth}{0.5pt}  
	\resizebox{\textwidth}{!}{%
		\scriptsize
		\begin{tabular}{@{}c*{9}{c}@{}}
			\toprule  
			\multirow{4}{*}{Method} & \multicolumn{3}{c}{Path-Cycle} & \multicolumn{3}{c}{Path-Fan} & \multicolumn{3}{c}{Complete-Star} \\
			
			\cmidrule(lr){2-4} \cmidrule(lr){5-7} \cmidrule(lr){8-10}
			& $\sigma$=0.5 & $\sigma$=1.0 & $\sigma$=1.5
			& $\sigma$=0.5 & $\sigma$=1.0 & $\sigma$=1.5 
			& $\sigma$=0.5 & $\sigma$=1.0& $\sigma$=1.5 \\
			\midrule  
			2D-GFRFT\textsubscript{UU}
			& \makecell{$10.667$ \\ $(1.019)$} & \makecell{$10.7134$ \\ $(1.018)$} & \makecell{$10.8146$ \\ $(-0.509)$}
			& \makecell{$5.2414$ \\ $(0.816)$} & \makecell{$5.3326$ \\ $(0.823)$} & \makecell{$5.7955$ \\ $(-0.514)$}
			& \makecell{$5.7303$ \\ $(0.051)$} & \makecell{$5.8434$ \\ $(1.006)$} & \makecell{$5.9098$ \\ $(1.017)$} \\
			2D-GBFRFT\textsubscript{UU}
			& \makecell{$\mathbf{10.4153}$ \\ $(0.902, 1.255)$} & \makecell{$\mathbf{10.6724}$ \\ $(-0.012, 1.323)$} & \makecell{$\mathbf{10.7414}$ \\ $(-0.013, 1.322)$}
			& \makecell{$\mathbf{5.1449}$ \\ $(0.893, 0.588)$} & \makecell{$\mathbf{5.2466}$ \\ $(0.909, 0.6)$} & \makecell{$\mathbf{5.6737}$ \\ $(0.141, 0.546)$}
			& \makecell{$\mathbf{5.4348}$ \\ $(0.691, 0.125)$} & \makecell{$\mathbf{5.5534}$ \\ $(0.681, 0.132)$} & \makecell{$\mathbf{5.6836}$ \\ $(0.672, 0.138)$} \\
			2D-GFRFT\textsubscript{UW}
			& \makecell{$10.6873$ \\ $(-0.062)$} & \makecell{$10.7327$ \\ $(-0.067)$} & \makecell{$10.7946$ \\ $(-0.063)$}
			& \makecell{$5.2461$ \\ $(-0.046)$} & \makecell{$5.3471$ \\ $(-0.044)$} & \makecell{$5.4657$ \\ $(-0.044)$}
			& \makecell{$5.6473$ \\ $(0.027)$} & \makecell{$5.7098$ \\ $(0.602)$} & \makecell{$5.7949$ \\ $(0.609)$} \\
			2D-GBFRFT\textsubscript{UW}
			& \makecell{$\mathbf{10.6579}$ \\ $(0.949, 0.036)$} & \makecell{$\mathbf{10.7025}$ \\ $(0.96, 0.025)$} & \makecell{$\mathbf{10.7186}$ \\ $(0.154, 0.26)$}
			& \makecell{$\mathbf{4.9362}$ \\ $(0.472, 0.604)$} & \makecell{$\mathbf{5.0709}$ \\ $(0.473, 0.601)$} & \makecell{$\mathbf{5.2337}$ \\ $(0.475, 0.6)$}
			& \makecell{$\mathbf{5.5376}$ \\ $(0.161, -0.004)$} & \makecell{$\mathbf{5.6292}$ \\ $(0.168, -0.005)$} & \makecell{$\mathbf{5.7348}$ \\ $(0.174, -0.01)$} \\
			2D-GFRFT\textsubscript{DU}
			& \makecell{$31.1218$ \\ $(-0.91)$} & \makecell{$31.1425$ \\ $(-0.909)$} & \makecell{$31.1746$ \\ $(-0.913)$}
			& \makecell{$18.735$ \\ $(0.496)$} & \makecell{$18.7735$ \\ $(0.5)$} & \makecell{$18.8319$ \\ $(0.506)$}
			& \makecell{$8.1761$ \\ $(0.693)$} & \makecell{$8.2317$ \\ $(0.69)$} & \makecell{$8.308$ \\ $(0.687)$} \\
			2D-GBFRFT\textsubscript{DU}
			& \makecell{$\mathbf{20.4298}$ \\ $(0.5, -0.602)$} & \makecell{$\mathbf{20.4683}$ \\ $(0.5, -0.6)$} & \makecell{$\mathbf{20.525}$ \\ $(0.5, -0.597)$}
			& \makecell{$\mathbf{7.8237}$ \\ $(1.259, 1.986)$} & \makecell{$\mathbf{7.9218}$ \\ $(1.244, 1.206)$} & \makecell{$\mathbf{8.1498}$ \\ $(1.331, 0.773)$}
			& \makecell{$\mathbf{8.1748}$ \\ $(0.723, 0.680)$} & \makecell{$\mathbf{8.2306}$ \\ $(0.719, 0.677)$} & \makecell{$\mathbf{8.307}$ \\ $(0.715, 0.673)$} \\
			2D-GFRFT\textsubscript{DW}
			& \makecell{$30.9344$ \\ $(-0.742)$} & \makecell{$30.9697$ \\ $(-0.704)$} & \makecell{$31.0065$ \\ $(-0.714)$}
			& \makecell{$19.1701$ \\ $(0.902)$} & \makecell{$19.2104$ \\ $(0.909)$} & \makecell{$19.2728$ \\ $(0.896)$}
			& \makecell{$8.0321$ \\ $(1.459)$} & \makecell{$8.0961$ \\ $(1.459)$} & \makecell{$8.1844$ \\ $(1.462)$} \\
			2D-GBFRFT\textsubscript{DW}
			& \makecell{$\mathbf{20.1816}$ \\ $(0.29, -0.793)$} & \makecell{$\mathbf{20.2207}$ \\ $(0.296, -0.792)$} & \makecell{$\mathbf{20.2798}$ \\ $(0.301, -0.788)$}
			& \makecell{$\mathbf{8.1930}$ \\ $(0.354, 0.125)$} & \makecell{$\mathbf{8.2758}$ \\ $(0.35, 0.125)$} & \makecell{$\mathbf{8.3865}$ \\ $(0.344, 0.125)$}
			& \makecell{$\mathbf{8.0072}$ \\ $(1.508, 0.631)$} & \makecell{$\mathbf{8.1029}$ \\ $(1.540, 0.758)$} & \makecell{$\mathbf{8.1791}$ \\ $(1.515, 1.375)$} \\
			\bottomrule  
		\end{tabular}
	}
	\\[0.5ex]
	\footnotesize \textbf{Note:} The subscripts indicate graph types: 
	UU = undirected unweighted, 
	UW = undirected weighted, 
	DU = directed unweighted, 
	DW = directed weighted.
\end{table*}

\begin{figure}[htbp]
    \centering
    \includegraphics[width=0.9\columnwidth]{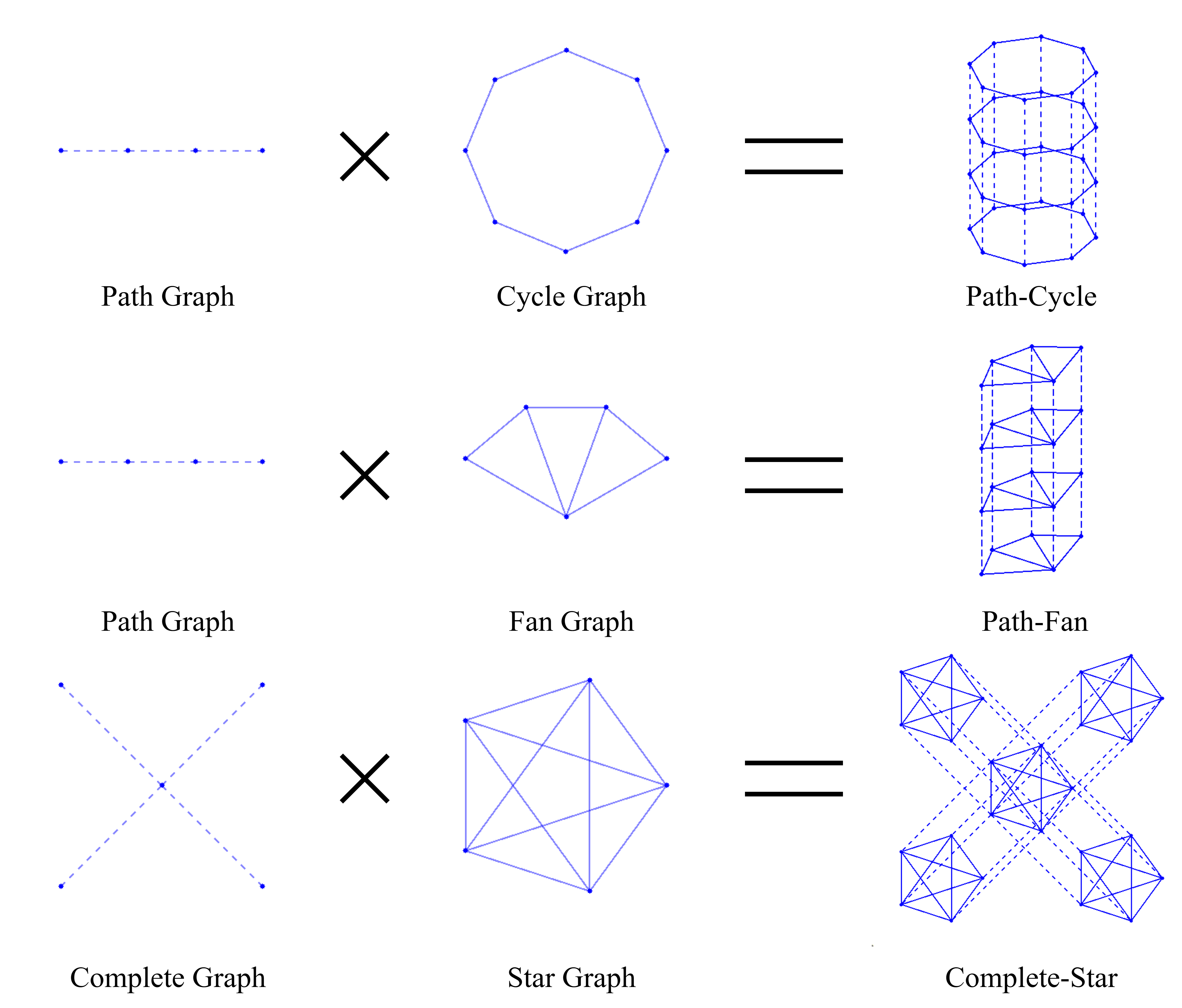}
    \caption{Cartesian product graphs of path-cycle, path-fan, and complete-star. Each row shows two factor graphs (left and middle) and their Cartesian product (right).}
    \label{fig:cartesian_product}
\end{figure}

We evaluate the proposed 2D-GBFRFT against the conventional 2D-GFRFT using synthetic signals defined on Cartesian product graphs. Three representative Cartesian products are illustrated in Fig.~\ref{fig:cartesian_product}: path-cycle, where a path graph $P_4$ and a cycle graph $C_8$ form a cylindrical grid; path-fan, constructed from a path graph $P_4$ and a fan graph $F_5$; and complete-star, formed by a complete graph $K_5$ and a star graph $S_5$. Each factor graph is generated according to its intended topology (directed or undirected, weighted or unweighted). The adjacency matrix is determined by the structure of the individual graphs, with edge directions and weights assigned based on the specific topology of each factor graph. The adjacency matrix of the Cartesian product is then obtained via the standard Cartesian product operation.
Motivated by the work in~\cite{42}, 
for a Cartesian product graph with $N_1N_2$ nodes, the signal autocorrelation matrix 
$\mathbf{C}\in\mathbb{R}^{N_1N_2\times N_1N_2}$ is defined as
\begin{equation}
\mathbf{C}_{ij} =
\begin{cases}
2, & \text{if } i = j,\\
1, & \text{if nodes $i$ and $j$ are adjacent},\\
0, & \text{otherwise}.
\end{cases}
\label{eq:autocorr}
\end{equation}
This formulation ensures that nonzero correlation exists only between directly connected nodes. Normalizing by the largest eigenvalue $\lambda_{\max}(\mathbf{C})$ yields $\mathbf{R}_{\mathbf{x}\mathbf{x}} = \mathbf{C}/\lambda_{\max}(\mathbf{C})$, and the average signal power is given by $\mathbb{E}\{\mathbf{x}^T\mathbf{x}\} = 2N_1N_2/\lambda_{\max}(\mathbf{C})$. Clean signals $\mathbf{x}$ are sampled from a zero-mean Gaussian process with covariance $\mathbf{R}_{\mathbf{x}\mathbf{x}}$, and noisy observations are obtained as $\mathbf{y}=\mathbf{x}+\mathbf{n}$, where $\mathbf{n}$ is zero-mean Gaussian noise with variance $\sigma^2\in\{0.5,1.0,1.5\}$.

Two parameter estimation strategies are considered. In grid search, the fractional order $\alpha$ of 2D-GFRFT is scanned over $[0,1]$ with step size $0.1$, while for 2D-GBFRFT the two independent orders $(\alpha_1,\alpha_2)$ are jointly scanned over $[0,1]\times[0,1]$ with the same step size. In gradient descent, the transform orders and diagonal filter coefficients are initialized uniformly in $[-1,1]$ and jointly optimized using the Adam optimizer with learning rate $0.03$ for $200$ iterations.

The results in Tables~\ref{tab:grid} and \ref{tab:grad} show that 2D-GBFRFT consistently outperforms 2D-GFRFT across all three Cartesian product graphs, with the performance gain being most significant when the factor graphs exhibit distinct spectral characteristics. This advantage arises from the ability of 2D-GBFRFT to assign independent fractional orders along each factor-graph dimension, enabling anisotropic spectral compression that enhances separability between signal and noise.

Experimental results show that both grid search and gradient descent achieve comparable denoising performance. Nevertheless, gradient descent offers several practical advantages: it avoids exhaustive enumeration, requires no predefined parameter range, and jointly optimizes the transform orders and filter coefficients within a unified framework. This allows better adaptation to the underlying signal and noise statistics while substantially reducing the computational cost. Considering both accuracy and efficiency, all subsequent experiments employ gradient descent for parameter estimation.

\section{Applications on Real-World Data}
\label{sec:realdata}
\subsection{Denoising of Time-Varying Graph Signals}

\begin{table*}[htbp]
\centering
\caption{DENOISING RESULTS (MSE) ON THE COVID DATASET}
\label{tab:COVID_denoising_results}
\resizebox{\textwidth}{!}{%
\scriptsize
\begin{tabular}{@{}c*{9}{c}@{}}
\toprule
\multirow{4}{*}{Method} & \multicolumn{3}{c}{3-NN} & \multicolumn{3}{c}{4-NN} & \multicolumn{3}{c}{5-NN} \\
\cmidrule(lr){2-4} \cmidrule(lr){5-7} \cmidrule(lr){8-10}
 & $\sigma$=0.6 & $\sigma$=0.9 & $\sigma$=1.2
 & $\sigma$=0.6 & $\sigma$=0.9 & $\sigma$=1.2
 & $\sigma$=0.6 & $\sigma$=0.9 & $\sigma$=1.2 \\
\midrule
\makecell{2D-GFRFT \\ $(\alpha)$}
& \makecell{$0.6424$ \\ $(0.307)$} & \makecell{$1.0626$ \\ $(0.263)$} & \makecell{$1.0749$ \\ $(-0.485)$}
& \makecell{$0.4901$ \\ $(0.285)$} & \makecell{$0.8104$ \\ $(0.298)$} & \makecell{$1.1075$ \\ $(0.788)$}
& \makecell{$0.3828$ \\ $(0.682)$} & \makecell{$0.5896$ \\ $(0.860)$} & \makecell{$0.7687$ \\ $(1.184)$} \\
\makecell{2D-GBFRFT \\ $(\alpha_1, \alpha_2)$}
& \makecell{\textbf{0.1079} \\ $(0.036, 0.999)$} & \makecell{\textbf{0.2160} \\ $(0.025, -0.999)$} & \makecell{\textbf{0.3547} \\ $(0.020, 1.000)$}
& \makecell{\textbf{0.1081} \\ $(-0.003, -1.000)$} & \makecell{$0.6872$ \\ $(0.963, -0.566)$} & \makecell{$0.9820$ \\ $(0.937, 0.564)$}
& \makecell{$0.4134$ \\ $(0.490, -0.306)$} & \makecell{\textbf{0.2183} \\ $(0.031, 0.997)$} & \makecell{\textbf{0.3541} \\ $(0.037, 0.997)$} \\
\makecell{JFRFT \\ $(\alpha, \beta)$}
& \makecell{$0.5216$ \\ $(0.853, 1.576)$} & \makecell{$0.8382$ \\ $(0.894, 1.543)$} & \makecell{$0.9399$ \\ $(1.135, -0.004)$}
& \makecell{$0.5071$ \\ $(0.791, 0.609)$} & \makecell{$0.8189$ \\ $(0.770, 0.644)$} & \makecell{$1.1102$ \\ $(0.774, -0.661)$}
& \makecell{$0.2853$ \\ $(1.049, 0.834)$} & \makecell{$0.4920$ \\ $(1.026, 0.990)$} & \makecell{$0.7108$ \\ $(1.015, 0.994)$} \\
\makecell{(2DGB+J)FRFT \\ $(\alpha, \beta, \lambda)$}
& \makecell{\textbf{0.1079} \\ $(0.036, 0.999, 0.000)$} & \makecell{\textbf{0.2160} \\ $(0.025, -0.999, 0.000)$} & \makecell{\textbf{0.3547} \\ $(0.020, 1.000, 0.000)$}
& \makecell{\textbf{0.1081} \\ $(-0.003, -1.000, 0.000)$} & \makecell{\textbf{0.2736} \\ $(-0.021, 1.017, 0.900)$} & \makecell{\textbf{0.4831} \\ $(-0.040, 0.999, 0.100)$}
& \makecell{\textbf{0.1213} \\ $(0.072, 1.027, 0.900)$} & \makecell{\textbf{0.2183} \\ $(0.031, 0.997, 0.000)$} & \makecell{\textbf{0.3541} \\ $(0.037, 0.997, 0.000)$} \\

\bottomrule
\end{tabular}
}
\end{table*}

\begin{table*}[htbp]
\centering
\caption{DENOISING RESULTS (MSE) ON THE SST DATASET}
\label{tab:SST_denoising_results}
\resizebox{\textwidth}{!}{
\scriptsize
\begin{tabular}{@{}c*{9}{c}@{}}
\toprule
\multirow{4}{*}{Method} & \multicolumn{3}{c}{3-NN} & \multicolumn{3}{c}{4-NN} & \multicolumn{3}{c}{5-NN} \\
\cmidrule(lr){2-4} \cmidrule(lr){5-7} \cmidrule(lr){8-10}
 & $\sigma$=0.6 & $\sigma$=0.9 & $\sigma$=1.2
 & $\sigma$=0.6 & $\sigma$=0.9 & $\sigma$=1.2
 & $\sigma$=0.6 & $\sigma$=0.9 & $\sigma$=1.2 \\
\midrule
\makecell{2D-GFRFT \\ $(\alpha)$}
& \makecell{$0.7200$ \\ $(0.698)$} & \makecell{$1.5089$ \\ $(0.684)$} & \makecell{$2.4722$ \\ $(0.668)$}
& \makecell{$0.7888$ \\ $(0.262)$} & \makecell{$1.6653$ \\ $(0.262)$} & \makecell{$2.7415$ \\ $(0.265)$}
& \makecell{$0.6310$ \\ $(0.783)$} & \makecell{$1.2809$ \\ $(0.792)$} & \makecell{$2.0307$ \\ $(0.820)$} \\
\makecell{2D-GBFRFT \\ $(\alpha_1, \alpha_2)$}
& \makecell{$\textbf{0.6579}$ \\ $(0.610, 0.713)$} & \makecell{$\textbf{1.3623}$ \\ $(0.572, 0.741)$} & \makecell{$\textbf{2.2080}$ \\ $(0.528, 0.768)$}
& \makecell{$\textbf{0.5273}$ \\ $(0.172, 0.900)$} & \makecell{$1.1355$ \\ $(0.174, 0.837)$} & \makecell{$1.7663$ \\ $(0.165, 1.004)$}
& \makecell{$0.6784$ \\ $(0.735, 0.279)$} & \makecell{$1.2509$ \\ $(0.753, 0.390)$} & \makecell{$1.9837$ \\ $(0.760, 0.484)$} \\
\makecell{JFRFT \\ $(\alpha, \beta)$}
& \makecell{$1.1687$ \\ $(0.750, 0.746)$} & \makecell{$1.8419$ \\ $(0.250, 0.659)$} & \makecell{$2.8968$ \\ $(0.240, 0.628)$}
& \makecell{$2.1764$ \\ $(0.430, 0.447)$} & \makecell{$1.3682$ \\ $(0.179, 0.751)$} & \makecell{$2.0061$ \\ $(0.185, 0.987)$}
& \makecell{$0.7092$ \\ $(0.841, 0.502)$} & \makecell{$0.9451$ \\ $(0.721, 0.629)$} & \makecell{$\textbf{1.4215}$ \\ $(0.671, 0.716)$} \\
\makecell{(2DGB+J)FRFT \\ $(\alpha,\beta,\lambda)$}
& \makecell{$\textbf{0.6579}$ \\ $(0.610, 0.713, 0.000)$} & \makecell{$\textbf{1.3623}$ \\ $(0.572, 0.741, 0.000)$} & \makecell{$\textbf{2.2080}$ \\ $(0.528, 0.768, 0.000)$}
& \makecell{$\textbf{0.5273}$ \\ $(0.172, 0.900, 0.000)$} & \makecell{$\textbf{0.9118}$ \\ $(1.011, 1.443, 0.300)$} & \makecell{$\textbf{1.6058}$ \\ $(1.025, 1.475, 0.200)$}
& \makecell{$\textbf{0.4183}$ \\ $(0.672, 0.845, 0.900)$} & \makecell{$\textbf{0.8950}$ \\ $(0.667, 0.822, 0.900)$} & \makecell{$\textbf{1.4215}$ \\ $(0.671, 0.716, 1.000)$} \\

\bottomrule
\end{tabular}
}
\end{table*}

\begin{table*}[htbp]
\centering
\caption{DENOISING RESULTS (MSE) ON THE PM-25 DATASET}
\label{tab:PM-25_denoising_results}
\resizebox{\textwidth}{!}{%
\scriptsize
\begin{tabular}{@{}c*{9}{c}@{}}
\toprule
\multirow{4}{*}{Method} & \multicolumn{3}{c}{3-NN} & \multicolumn{3}{c}{4-NN} & \multicolumn{3}{c}{5-NN} \\
\cmidrule(lr){2-4} \cmidrule(lr){5-7} \cmidrule(lr){8-10}
 & $\sigma$=0.6 & $\sigma$=0.9 & $\sigma$=1.2
 & $\sigma$=0.6 & $\sigma$=0.9 & $\sigma$=1.2
 & $\sigma$=0.6 & $\sigma$=0.9 & $\sigma$=1.2 \\
\midrule
\makecell{2D-GFRFT \\ $(\alpha)$}
& \makecell{$0.6636$ \\ $(0.813)$} & \makecell{$1.2474$ \\ $(0.804)$} & \makecell{$1.8847$ \\ $(0.801)$}
& \makecell{$0.9222$ \\ $(0.584)$} & \makecell{$1.7594$ \\ $(0.577)$} & \makecell{$2.1018$ \\ $(1.014)$}
& \makecell{$0.6605$ \\ $(0.315)$} & \makecell{$1.1877$ \\ $(0.318)$} & \makecell{$1.7555$ \\ $(0.320)$} \\
\makecell{2D-GBFRFT\\ $(\alpha_1, \alpha_2)$}
& \makecell{$0.6449$ \\ $(0.800, 0.646)$} & \makecell{$1.2081$ \\ $(0.765, 0.606)$} & \makecell{$1.7727$ \\ $(0.957, 0.738)$}
& \makecell{$0.7685$ \\ $(0.614, -0.233)$} & \makecell{$1.5059$ \\ $(0.618, -0.273)$} & \makecell{$2.2999$ \\ $(0.624, 0.275)$}
& \makecell{$0.6523$ \\ $(0.329, 0.280)$} & \makecell{$1.1854$ \\ $(0.322, 0.298)$} & \makecell{$1.7549$ \\ $(0.322, 0.311)$} \\
\makecell{JFRFT\\ $(\alpha, \beta)$}
& \makecell{\textbf{0.5326} \\ $(0.643, 0.480)$} & \makecell{\textbf{0.8955} \\ $(0.928, 0.594)$} & \makecell{\textbf{1.3698} \\ $(0.822, 0.559)$}
& \makecell{$0.4295$ \\ $(0.926, 0.532)$} & \makecell{\textbf{0.8629} \\ $(0.977, 0.567)$} & \makecell{\textbf{1.3931} \\ $(0.974, 0.564)$}
& \makecell{\textbf{0.4497}\\ $(0.480, 0.541)$} & \makecell{\textbf{0.9242} \\ $(0.490, 0.539)$} & \makecell{\textbf{1.4784} \\ $(0.505, 0.535)$} \\
\makecell{(2DGB+J)FRFT \\ $(\alpha, \beta, \lambda)$}
& \makecell{\textbf{0.5326} \\ $(0.643, 0.480, 1.000)$} & \makecell{\textbf{0.8955} \\ $(0.928, 0.594, 1.000)$} & \makecell{\textbf{1.3698} \\ $(0.822, 0.559, 1.000)$}
& \makecell{\textbf{0.4253} \\ $(1.000, 0.618, 0.900)$} & \makecell{\textbf{0.8629} \\ $(0.977, 0.567, 1.000)$} & \makecell{\textbf{1.3931} \\ $(0.974, 0.564, 1.000)$}
& \makecell{\textbf{0.4497} \\ $(0.480, 0.541, 1.000)$} & \makecell{\textbf{0.9242} \\ $(0.490, 0.539, 1.000)$} & \makecell{\textbf{1.4784} \\ $(0.505, 0.535, 1.000)$} \\
\bottomrule
\end{tabular}
}
\end{table*}

In~\cite{1}, time-varying graph signals are modeled by representing the temporal dimension as a path graph and the spatial relationships as another subgraph, forming a Cartesian product graph. This approach enables direct comparison with JFRFT for the same problem. We evaluate 2D-GFRFT, 2D-GBFRFT, and JFRFT on real-world time-varying graph signals. In contrast to the synthetic experiments in Section~\ref{sec:synthetic}, which use artificially constructed Cartesian product graphs, this section examines real-world time-varying data. Within the Cartesian product framework, the temporal dimension is modeled by a path graph, while the spatial dimension is represented by an undirected $k$-nearest neighbor ($k$-NN) graph. The $k$-NN graph is constructed using node coordinates, with $k \in \{3,4,5\}$.

The datasets include Sea Surface Temperature (SST), Particulate Matter 2.5 (PM-25), and COVID-19 global (COVID) case time series, each represented as a temporal graph signal with 20 spatial nodes and 3 time steps. To enable quantitative evaluation, Gaussian noise is added to the standardized signals. The noise variances are selected from $\sigma^2 \in \{0.6, 0.9, 1.2\}$. The performance metric used is MSE, computed relative to the clean reference signals.

Experimental results on real temporal graph signals reveal that the relative performance of JFRFT and 2D-GBFRFT depends on the characteristics of the dataset. On the COVID time series (Table~\ref{tab:COVID_denoising_results}), 2D-GBFRFT achieves superior denoising performance, benefiting from its ability to apply independent fractional orders along temporal and spatial dimensions. In contrast, on the PM-25 dataset (Table~\ref{tab:PM-25_denoising_results}), JFRFT performs better, as the signals are dominated by temporal smoothness and are well captured by a fractional Fourier basis along the temporal path. Motivated by this complementary behavior, we introduce a hybrid transform, denoted (2DGB+J)FRFT, which interpolates between JFRFT and 2D-GBFRFT through a tunable hyperparameter $\lambda \in [0,1]$. The transform is defined as
\begin{equation}
	\mathbf{X}_{\mathrm f}
	= \mathbf{F}^{\alpha}_{G_1}\,\mathbf{X}\,
	\big[\lambda\,\mathbf{F}^{\beta}+(1-\lambda)\,\mathbf{F}^{\beta}_{G_2}\big]^{\top},
	\label{eq:hybrid-frft}
\end{equation}
where $\mathbf{F}^{\alpha}_{G_1}$ is the spatial graph fractional basis, 
$\mathbf{F}^{\beta}_{G_2}$ is the temporal path-graph fractional basis, 
and $\mathbf{F}^{\beta}$ is the standard discrete FRFT basis. The vectorized operator follows the same column-wise vectorization convention as in 2D-GBFRFT.
The interpolation parameter $\lambda$ is selected by grid search with step size $0.1$. 
When $\lambda=1$, the transform reduces to JFRFT; 
when $\lambda=0$, it coincides with the bi-factorized form of 2D-GBFRFT.

For fairness, all methods adopt the same gradient descent scheme as in Section~\ref{sec:synthetic}, but with different hyperparameters adapted to real data: learning rate $0.1$, 200 iterations, and the fractional orders were uniformly initialized to $0.5$, with the diagonal filter initialized as identity. This ensures a consistent training environment across methods, while the separate grid search for $\lambda$ avoids conflating interpolation with spectral order optimization.

The filtering procedure of (2DGB+J)FRFT under gradient descent follows the same theoretical principles as the 2D-GBFRFT case described earlier, with the main difference being the additional interpolation parameter $\lambda$. The optimization consists of an outer grid search over $\lambda$ and inner-loop gradient descent updates for the transform orders and diagonal filter, as summarized in Algorithm~\ref{alg:hybrid-frft-gd}.

\begin{algorithm}[t]
\caption{(2DGB+J)FRFT with Outer $\lambda$ Grid and Inner Gradient Descent}
\label{alg:hybrid-frft-gd}
\begin{algorithmic}[1]
\REQUIRE $\mathbf{y}=(\mathbf{G}_{G_2}^{\top}\!\otimes\!\mathbf{G}_{G_1})\mathbf{x}+\mathbf{n}$, learning rate $\gamma$, epochs $L$, grid $\Lambda=\{0,0.1,\ldots,1\}$
\ENSURE Selected $\lambda^\ast$, trained orders $(\alpha^\ast,\beta^\ast)$, diagonal filter $\mathbf{H}_{2D}^\ast$, final MSE

\FOR{$\lambda \in \Lambda$}
  \STATE Initialize $\alpha=0.5$, $\beta=0.5$, $\mathbf{H}_{2D}=\mathbf{I}$
  \FOR{$t=1$ to $L$}
    \STATE \textbf{Step 1: Operator construction}
    \STATE Build $\mathbf{T}_\lambda(\beta)=\lambda\,\mathbf{F}^{\beta}+(1-\lambda)\,\mathbf{F}_{G_2}^{\beta}$
    \STATE Build $\mathbf{F}_{2D}^{(\alpha,\beta,\lambda)}=\mathbf{T}_\lambda(\beta)\otimes\mathbf{F}_{G_1}^{\alpha}$
    \STATE \textbf{Step 2: Forward propagation}
    \STATE $\hat{\mathbf{x}}=(\mathbf{F}_{2D}^{(\alpha,\beta,\lambda)})^{-1}\mathbf{H}_{2D}\,\mathbf{F}_{2D}^{(\alpha,\beta,\lambda)}\,\mathbf{y}$
    \STATE \textbf{Step 3: Loss calculation and update}
    \STATE $\mathcal{L}=\|\hat{\mathbf{x}}-\mathbf{x}\|_2^2$
    \STATE Update $(\alpha,\beta,\mathbf{H}_{2D})$ by gradient descent
  \ENDFOR
  \STATE Compute $\mathrm{MSE}(\lambda)=\|\hat{\mathbf{x}}-\mathbf{x}\|_2^2$
  \IF{first $\lambda$ \textbf{or} $\mathrm{MSE}(\lambda)$ is smaller than current best}
    \STATE Store $\lambda^\ast=\lambda$, $\alpha^\ast=\alpha$, $\beta^\ast=\beta$, $\mathbf{H}_{2D}^\ast=\mathbf{H}_{2D}$, $\mathrm{MSE}^\ast=\mathrm{MSE}(\lambda)$
  \ENDIF
\ENDFOR
\STATE \textbf{Return} $\lambda^\ast$, $(\alpha^\ast,\beta^\ast)$, $\mathbf{H}_{2D}^\ast$, $\mathrm{MSE}^\ast$
\end{algorithmic}
\end{algorithm}

As shown in Tables~\ref{tab:COVID_denoising_results}, \ref{tab:SST_denoising_results}, and \ref{tab:PM-25_denoising_results}, 2D-GBFRFT outperforms 2D-GFRFT in the majority of results across the datasets, validating the flexibility of applying independent fractional orders to both spatial and temporal dimensions, although certain cases exhibit suboptimal performance due to the possibility of gradient descent being trapped in local minima. The comparative results with JFRFT depend on the spatio-temporal characteristics of the data.

Furthermore, by introducing the hybrid (2DGB+J)FRFT, which interpolates between JFRFT and 2D-GBFRFT via a tunable hyperparameter $\lambda$, the method adapts to the strengths of both approaches. The hybrid transform consistently achieves the best denoising performance across all datasets, irrespective of the dataset's characteristics, demonstrating the versatility and robustness of (2DGB+J)FRFT in balancing spatial and temporal representations for real-world time-varying graph signals.

\begin{figure*}[t!]
\centering

\setlength{\tabcolsep}{2pt} 
\renewcommand{\arraystretch}{0}
\begin{tabular}{cccccc}
\includegraphics[width=0.145\textwidth]{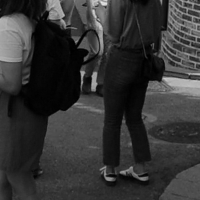} &
\includegraphics[width=0.145\textwidth]{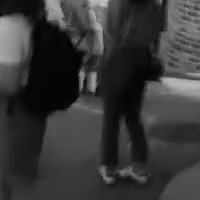} &
\includegraphics[width=0.145\textwidth]{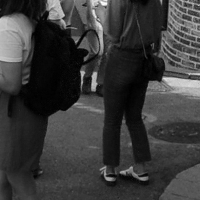} &
\includegraphics[width=0.145\textwidth]{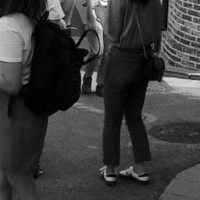} &
\includegraphics[width=0.145\textwidth]{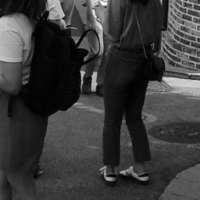} &
\includegraphics[width=0.145\textwidth]{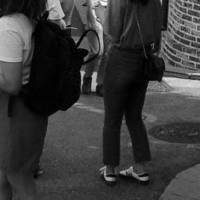} 
\\[1ex]
\includegraphics[width=0.145\textwidth]{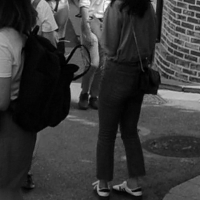} &
\includegraphics[width=0.145\textwidth]{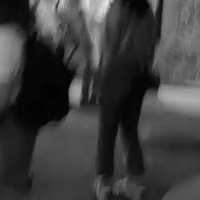} &
\includegraphics[width=0.145\textwidth]{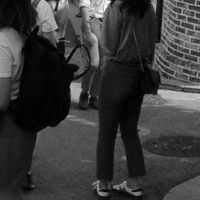} &
\includegraphics[width=0.145\textwidth]{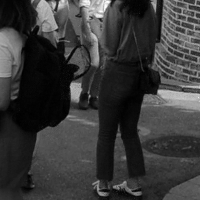} &
\includegraphics[width=0.145\textwidth]{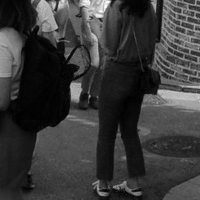} &
\includegraphics[width=0.145\textwidth]{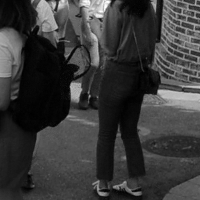} 
\\[1ex]
\includegraphics[width=0.145\textwidth]{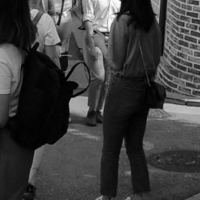} &
\includegraphics[width=0.145\textwidth]{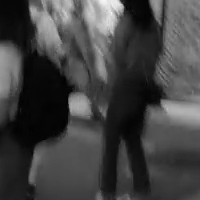} &
\includegraphics[width=0.145\textwidth]{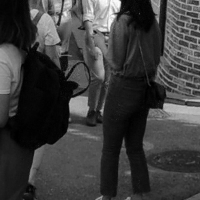} &
\includegraphics[width=0.145\textwidth]{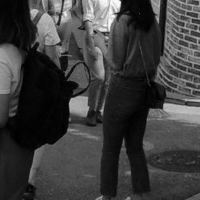} &
\includegraphics[width=0.145\textwidth]{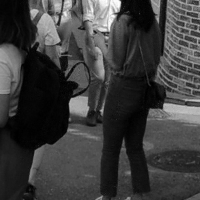} &
\includegraphics[width=0.145\textwidth]{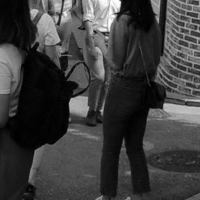}
\\[0.8ex]
Original & Blurred & 2D-GFRFT & 2D-GBFRFT & JFRFT & (2DGB+J)FRFT
\end{tabular}

\vspace{2pt} 

\setlength{\tabcolsep}{1pt}
\renewcommand{\arraystretch}{0}
\begin{tabular}{cccc}
\includegraphics[width=0.18\textwidth]{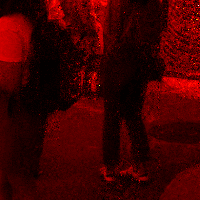} &
\includegraphics[width=0.18\textwidth]{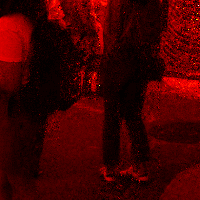} &
\includegraphics[width=0.18\textwidth]{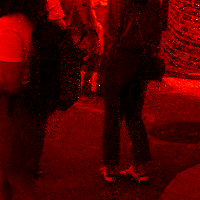} &
\includegraphics[width=0.18\textwidth]{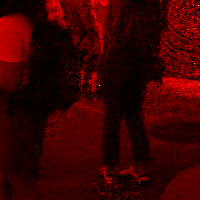} 
\\[0.8ex]
\includegraphics[width=0.18\textwidth]{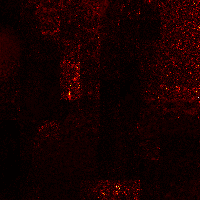} &
\includegraphics[width=0.18\textwidth]{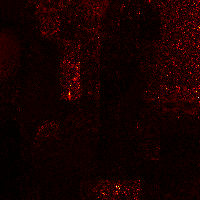} &
\includegraphics[width=0.18\textwidth]{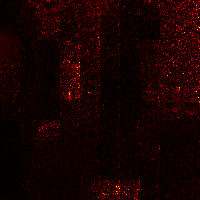} &
\includegraphics[width=0.18\textwidth]{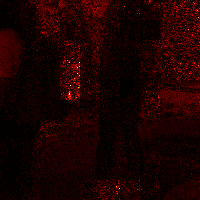} 
\\[0.8ex]
\includegraphics[width=0.18\textwidth]{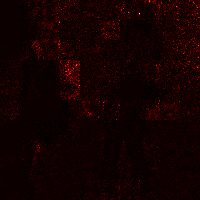} &
\includegraphics[width=0.18\textwidth]{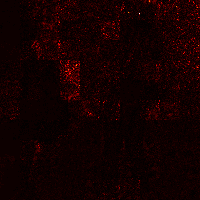} &
\includegraphics[width=0.18\textwidth]{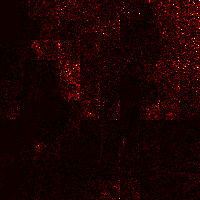} &
\includegraphics[width=0.18\textwidth]{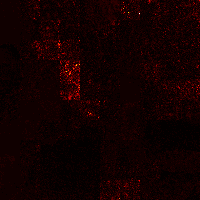} 
\\[1ex]
\small 2D-GFRFT & \small 2D-GBFRFT & \small JFRFT & \small (2DGB+J)FRFT
\end{tabular}

\caption{Recovering results and heatmap error visualization on the REDSA dataset. 
Top: restored frames under different transforms. 
Bottom: corresponding error heatmaps.}
\label{fig:REDSA_results_combined}
\end{figure*}

\begin{figure*}[t!]
\centering

\setlength{\tabcolsep}{2pt}
\renewcommand{\arraystretch}{0}
\begin{tabular}{cccccc}
\includegraphics[width=0.145\textwidth]{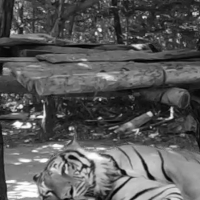} &
\includegraphics[width=0.145\textwidth]{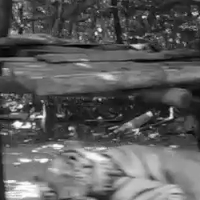} &
\includegraphics[width=0.145\textwidth]{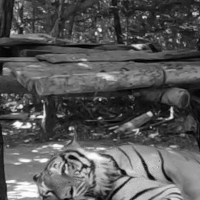} &
\includegraphics[width=0.145\textwidth]{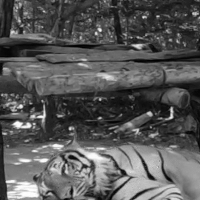} &
\includegraphics[width=0.145\textwidth]{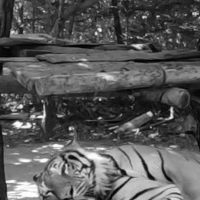} &
\includegraphics[width=0.145\textwidth]{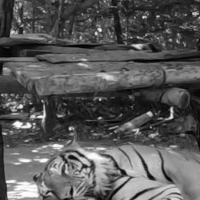} 
\\[1ex]
\includegraphics[width=0.145\textwidth]{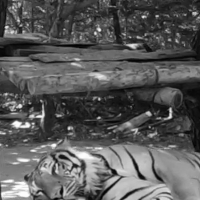} &
\includegraphics[width=0.145\textwidth]{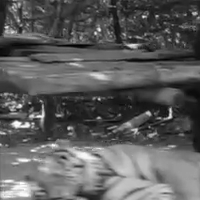} &
\includegraphics[width=0.145\textwidth]{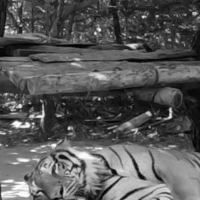} &
\includegraphics[width=0.145\textwidth]{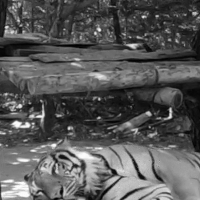} &
\includegraphics[width=0.145\textwidth]{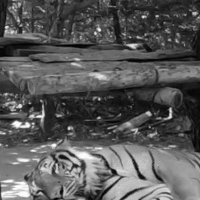} &
\includegraphics[width=0.145\textwidth]{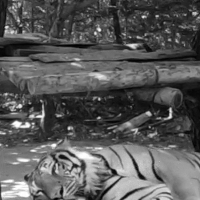} 
\\[1ex]
\includegraphics[width=0.145\textwidth]{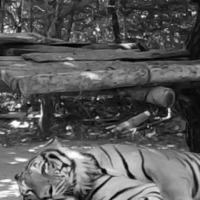} &
\includegraphics[width=0.145\textwidth]{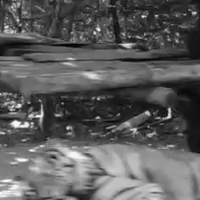} &
\includegraphics[width=0.145\textwidth]{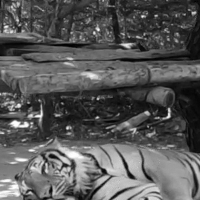} &
\includegraphics[width=0.145\textwidth]{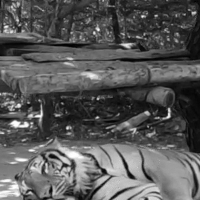} &
\includegraphics[width=0.145\textwidth]{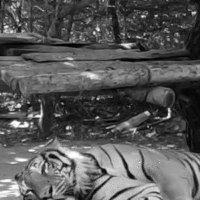} &
\includegraphics[width=0.145\textwidth]{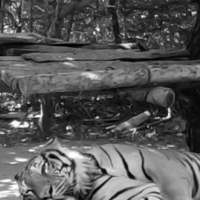}
\\[0.8ex]
Original & Blurred & 2D-GFRFT & 2D-GBFRFT & JFRFT & (2DGB+J)FRFT
\end{tabular}

\vspace{2pt} 

\setlength{\tabcolsep}{1pt}
\renewcommand{\arraystretch}{0}
\begin{tabular}{cccc}
\includegraphics[width=0.18\textwidth]{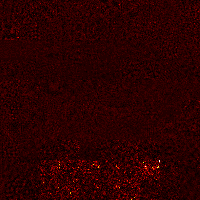} &
\includegraphics[width=0.18\textwidth]{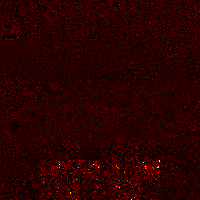} &
\includegraphics[width=0.18\textwidth]{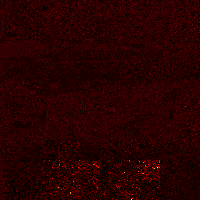} &
\includegraphics[width=0.18\textwidth]{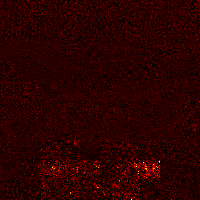} 
\\[0.8ex]
\includegraphics[width=0.18\textwidth]{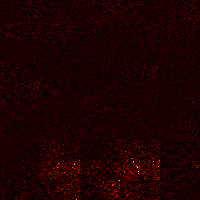} &
\includegraphics[width=0.18\textwidth]{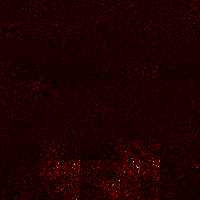} &
\includegraphics[width=0.18\textwidth]{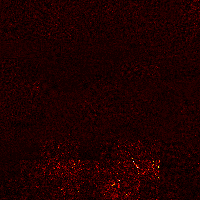} &
\includegraphics[width=0.18\textwidth]{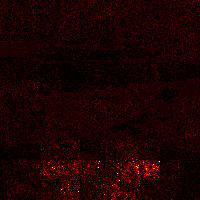} 
\\[0.8ex]
\includegraphics[width=0.18\textwidth]{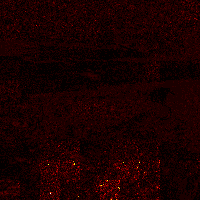} &
\includegraphics[width=0.18\textwidth]{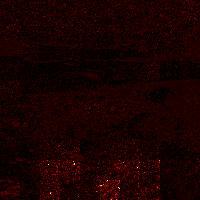} &
\includegraphics[width=0.18\textwidth]{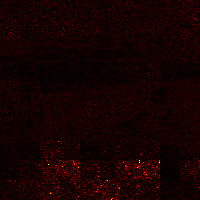} &
\includegraphics[width=0.18\textwidth]{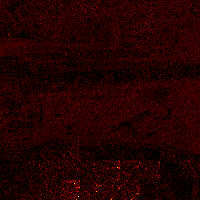} 
\\[1ex]
\small 2D-GFRFT & \small 2D-GBFRFT & \small JFRFT & \small (2DGB+J)FRFT
\end{tabular}

\caption{Recovering results and error heatmaps on the REDSB dataset. 
Top: restored frames under different transforms. 
Bottom: corresponding error heatmaps.}
\label{fig:REDSB_results_combined}
\end{figure*}

\subsection{Deblurring of Dynamic Images}  
We further evaluate the proposed 2D-GBFRFT for dynamic image deblurring, comparing its performance with 2D-GFRFT and JFRFT. The experiments are conducted on the REDS dataset, using two real-world scenes. For each scene, three consecutive frames are extracted, resized to $200 \times 200$ pixels, and divided into $100$ non-overlapping patches of size $20 \times 20$ pixels. Each set of three patches across consecutive frames is modeled as a spatio-temporal graph signal, where the grayscale pixel intensities serve as the signal values. The spatial graph is constructed as a $4$-nearest neighbor graph based on pixel locations, and the temporal dimension is represented by a $3$-node path graph.

Each $20 \times 20$ patch is treated as an independent training instance. The fractional orders and diagonal filter coefficients are optimized in a patch-wise manner, and the restored patches are reassembled to reconstruct the sequence. Adaptive filtering is applied under the respective transforms (2D-GFRFT, JFRFT, and (2DGB+J)FRFT), with MSE used as the loss function during training. Optimization is performed using the Adam optimizer with a learning rate of $7 \times 10^{-3}$ for $120$ epochs, with initial values for the orders set to $0.8$, and the filter matrix initialized as an identity matrix.  

For each frame \( t \), we compute the MSE, Peak Signal-to-Noise Ratio (PSNR), and Structural Similarity Index Measure (SSIM) as follows:
\begin{align}
	MSE_t &= \frac{1}{H W} \sum_{i,j} \left(I_t(i,j) - \hat{I}_t(i,j)\right)^2, \\
	PSNR_t &= 10 \log_{10}\!\left(\frac{\mathrm{MAX}^2}{MSE_t}\right), \\
	SSIM_t &= \frac{(2\mu_{I_t}\mu_{\hat{I}_t}+C_1)(2\sigma_{I_t\hat{I}_t}+C_2)}{(\mu_{I_t}^2+\mu_{\hat{I}_t}^2+C_1)(\sigma_{I_t}^2+\sigma_{\hat{I}_t}^2+C_2)},
\end{align}
where \( I_t \) and \( \hat{I}_t \) represent the ground-truth and reconstructed frames at time \( t \), respectively. Here, \( t \) denotes the time index or the frame number in the sequence of dynamic images. \( H \) and \( W \) are the height and width of the image, respectively, and are used to normalize the MSE computation. \( \mu_{I_t} \) and \( \mu_{\hat{I}_t} \) denote the means of \( I_t \) and \( \hat{I}_t \), while \( \sigma_{I_t} \) and \( \sigma_{\hat{I}_t} \) represent their standard deviations. The constants \( C_1 \) and \( C_2 \) are small values added to stabilize the division (typically set to \( 10^{-4} \) or \( 10^{-3} \)). \( \mathrm{MAX} \) represents the maximum possible pixel value, which is 255 for grayscale images. SSIM is computed using an $11 \times 11$ Gaussian window with a standard deviation of $1.5$. The metrics are further averaged over $T=3$ frames to obtain $MSE_{\text{avg}}$, $PSNR_{\text{avg}}$, and $SSIM_{\text{avg}}$.

\begin{table}[htbp]
\centering 
\caption{Comparison of Deblurring Performance (MSE/PSNR/SSIM) on the REDSA and REDSB Datasets}
\label{tab:deblurring_results_single}
\resizebox{\columnwidth}{!}{ 
\begin{tabular}{cccccc}
\toprule
Dataset & Method & Frame & MSE & PSNR & SSIM \\
\midrule
\multirow{12}{*}{REDSA}
& \multirow{3}{*}{2D-GFRFT} & Frame1 & $62.5371$ & $30.17$ & $0.9745$ \\
& & Frame2 & $7.9011$ & $39.15$ & $0.9840$ \\
& & Frame3 & $6.2759$ & $40.15$ & $0.9865$ \\
\cmidrule(lr){2-6}
& \multirow{3}{*}{2D-GBFRFT} & Frame1 & $60.0736$ & $30.34$ & $0.9758$ \\
& & Frame2 & $7.4058$ & $39.44$ & $0.9841$ \\
& & Frame3 & $5.8253$ & $40.48$ & $0.9869$ \\
\cmidrule(lr){2-6}
& \multirow{3}{*}{JFRFT} & Frame1 & $\textbf{54.5155}$ & $\textbf{30.77}$ & $\textbf{0.9823}$ \\
& & Frame2 & $8.6829$ & $38.74$ & $0.9825$ \\
& & Frame3 & $8.1092$ & $39.04$ & $0.9825$ \\
\cmidrule(lr){2-6}
& \multirow{3}{*}{(2DGB+J)FRFT} & Frame1 & $56.1140$ & $30.64$ & $0.9764$ \\
& & Frame2 & $\textbf{6.1070}$ & $\textbf{40.27}$ & $\textbf{0.9891}$ \\
& & Frame3 & $\textbf{3.4329}$ & $\textbf{42.77}$ & $\textbf{0.9891}$ \\
\midrule
\multirow{12}{*}{REDSB}
& \multirow{3}{*}{2D-GFRFT} 
& Frame1 & $8.6478$ & $38.76$ & $0.9931$ \\
& & Frame2 & $5.2380$ & $40.94$ & $0.9944$ \\
& & Frame3 & $2.1587$ & $44.79$ & $0.9967$ \\
\cmidrule(lr){2-6}
& \multirow{3}{*}{2D-GBFRFT} 
& Frame1 & $7.5175$ & $39.37$ & $0.9935$ \\
& & Frame2 & $5.1547$ & $41.01$ & $0.9945$ \\
& & Frame3 & $2.1082$ & $44.89$ & $0.9968$ \\
\cmidrule(lr){2-6}
& \multirow{3}{*}{JFRFT} & Frame1 & $\textbf{3.5065}$ & $\textbf{42.68}$ & $\textbf{0.9971}$ \\
& & Frame2 & $5.4131$ & $40.80$ & $0.9940$ \\
& & Frame3 & $4.8039$ & $41.31$ & $0.9929$ \\
\cmidrule(lr){2-6}
& \multirow{3}{*}{(2DGB+J)FRFT}
& Frame1 & $5.4029$ & $40.80$ & $0.9951$ \\
& & Frame2 & $\textbf{2.3247}$ & $\textbf{44.47}$ & $\textbf{0.9964}$ \\
& & Frame3 & $\textbf{1.8241}$ & $\textbf{45.52}$ & $\textbf{0.9980}$ \\
\bottomrule
\end{tabular}}
\end{table}

\begin{table}[htbp]
\centering 
\caption{Deblurring Performance in Terms of $MSE_{\text{avg}}$, $PSNR_{\text{avg}}$, and $SSIM_{\text{avg}}$ on the REDSA and REDSB Datasets}
\label{tab:deblurring_results_avg}
\resizebox{\columnwidth}{!}{ 
\begin{tabular}{ccccc}
\toprule
Dataset & Method & $MSE_{\text{avg}}$ & $PSNR_{\text{avg}}$ & $SSIM_{\text{avg}}$ \\
\midrule
\multirow{4}{*}{REDSA}
& 2D-GFRFT      & $25.5713$ & $36.49$ & $0.9817$ \\
& 2D-GBFRFT     & $24.4349$ & $36.75$ & $0.9823$ \\
& JFRFT        & $23.7692$ & $36.18$ & $0.9825$ \\
& (2DGB+J)FRFT & $\textbf{21.8846}$ & $\textbf{37.90}$ & $\textbf{0.9829}$ \\

\midrule
\multirow{4}{*}{REDSB}
& 2D-GFRFT      & $5.3482$ & $41.50$ & $0.9948$ \\
& 2D-GBFRFT    & $4.9268$ & $41.76$ & $0.9950$ \\
& JFRFT        & $4.5745$ & $41.60$ & $0.9947$ \\
& (2DGB+J)FRFT & $\textbf{3.1839}$ & $\textbf{43.60}$ & $\textbf{0.9965}$ \\

\bottomrule
\end{tabular}}
\end{table}

Quantitative results are presented in Tables~\ref{tab:deblurring_results_single} and \ref{tab:deblurring_results_avg}. Across both the REDSA and REDSB datasets, 2D-GBFRFT consistently outperforms 2D-GFRFT, highlighting the advantages of independent fractional orders along the spatial and temporal dimensions. The proposed (2DGB+J)FRFT leads to further improvements, with notable performance gains observed in the second and third frames of both REDSA and REDSB, while consistently achieving the best results on the average metrics.

Figs.~\ref{fig:REDSA_results_combined} and \ref{fig:REDSB_results_combined} provide additional insight into the results. In the restored frames, 2D-GBFRFT produces sharper results compared to 2D-GFRFT, while (2DGB+J)FRFT further enhances fine details. The error heatmaps offer more evidence of the improvements: compared to 2D-GFRFT, 2D-GBFRFT reduces error intensity around edges and textured regions, although residual distortions remain. In contrast, (2DGB+J)FRFT produces the clearest maps with the lowest error concentrations, significantly reducing both edge distortions and background noise. These visual and quantitative results together demonstrate that 2D-GBFRFT consistently outperforms 2D-GFRFT, and (2DGB+J)FRFT achieves the best performance across all average metrics.

\section{Conclusion}
\label{sec:conclusion}

This study presented 2D-GBFRFT, an innovative approach to graph signal processing that surpasses the conventional 2D-GFRFT in multiple experimental contexts. The experimental findings indicated that 2D-GBFRFT significantly improves denoising efficacy, especially for time-varying graph signals exhibiting significant spatial heterogeneity. In comparison to 2D-GFRFT, 2D-GBFRFT consistently produces enhanced denoising outcomes across various datasets, confirming the efficacy of independently adjusting fractional orders for both spatial and temporal dimensions.

To utilize the complementary advantages of 2D-GBFRFT and JFRFT, we proposed a hybrid transform, referred to as (2DGB+J)FRFT, which interpolates between the two techniques through a variable hyperparameter $\lambda$. Our findings indicate that (2DGB+J)FRFT consistently delivers superior performance across all datasets, achieving optimal denoising results regardless of the data's specific attributes. This underscores the flexibility and resilience of the hybrid transform in harmonizing temporal and spatial representations while consistently delivering exceptional denoising efficacy.

Despite its strengths, the proposed method has certain limitations. The interpolation parameter \( \lambda \) is restricted to the range \([0, 1]\), potentially limiting its flexibility in certain applications. Future research should explore extending the range of \( \lambda \) or optimizing it as a hyperparameter. Furthermore, this study primarily focuses on time-varying signals, especially those with subgraphs treated as path graphs; however, the application of the method to real-world Cartesian product graphs, which are devoid of path graphs, remains unexplored.

In summary, the 2D-GBFRFT and hybrid (2DGB+J)FRFT transforms offer powerful tools for denoising Cartesian product graph signals, particularly in the context of complex spatial heterogeneity. Future research will focus on broadening the scope of these methods, optimizing hyperparameters, and refining optimization strategies to enhance their applicability and performance.

\appendices
\section{Proof of Theorem 1}
The optimal filter coefficients are obtained by solving
\begin{equation}
	h_m^{\mathrm{opt}} = \arg \min_{h_m} \mathbb{E} \left\{
	\left\| \sum_{m=0}^{N_1 N_2 - 1} h_m \mathbf{W}_m \mathbf{y} - \mathbf{x} \right\|_2^2
	\right\}.
\end{equation}
Let \( \mathbf{S} = [\mathbf{s}_0, \mathbf{s}_1, \cdots, \mathbf{s}_{N_1 N_2 - 1}] \) with \( \mathbf{s}_k = \mathbf{W}_k \mathbf{y} \). This leads to the normal equations \( \mathbf{T} \mathbf{h}^{\mathrm{opt}} = \mathbf{q} \), where \( \mathbf{T} = \mathbb{E}\{\mathbf{S}^{\mathrm{H}} \mathbf{S}\} \) and \( \mathbf{q} = \mathbb{E}\{\mathbf{S}^{\mathrm{H}} \mathbf{x}\} \). The detailed expressions for \( \mathbf{T}_{m,n} \) and \( \mathbf{q}_m \) are:
\begin{IEEEeqnarray}{rCl}
	\mathbf{T}_{m,n} &=& 
	\mathrm{Tr}\!\left\{
	\mathbf{G}_{2D}^{\mathrm H}\mathbf{W}_m^{\mathrm H}\mathbf{W}_n
	\mathbf{G}_{2D}\,\mathbb{E}[\mathbf{x}\mathbf{x}^{\mathrm H}]
	\right\} \nonumber\\
	&&+ \mathrm{Tr}\!\left\{
	\mathbf{W}_m^{\mathrm H}\mathbf{W}_n\,\mathbb{E}[\mathbf{n}\mathbf{n}^{\mathrm H}]
	\right\} \nonumber\\
	&&+ \mathrm{Tr}\!\left\{
	\mathbf{G}_{2D}^{\mathrm H}\mathbf{W}_m^{\mathrm H}\mathbf{W}_n\,\mathbb{E}[\mathbf{n}\mathbf{x}^{\mathrm H}]
	\right\} \nonumber\\
	&&+ \mathrm{Tr}\!\left\{
	\mathbf{W}_m^{\mathrm H}\mathbf{W}_n\mathbf{G}_{2D}\,\mathbb{E}[\mathbf{x}\mathbf{n}^{\mathrm H}]
	\right\},
	\label{eq:T_entries}
\end{IEEEeqnarray}
\begin{IEEEeqnarray}{rCl}
	\mathbf{q}_m &=& 
	\mathrm{Tr}\!\left\{
	\mathbf{G}_{2D}^{\mathrm H}\mathbf{W}_m\,\mathbb{E}[\mathbf{x}\mathbf{x}^{\mathrm H}]
	\right\} \nonumber\\
	&&+ \mathrm{Tr}\!\left\{
	\mathbf{W}_m^{\mathrm H}\,\mathbb{E}[\mathbf{x}\mathbf{n}^{\mathrm H}]
	\right\},
	\label{eq:q_entries}
\end{IEEEeqnarray}
with \( \mathbf{W}_m = \mathbf{w}_m \widetilde{\mathbf{w}}_m^\top \), where \( \mathbf{w}_m^\top \) is the \( m \)-th row of \( \mathbf{F}_{2D}^{(\alpha_1, \alpha_2)} \) and \( \widetilde{\mathbf{w}}_m \) is the \( m \)-th column of \( \mathbf{F}_{2D}^{(-\alpha_1, -\alpha_2)} \).

\section{Proof of Theorem 2} 
In~\cite{29}, 
the differentiability can be derived via the matrix exponential parameterization.
\begin{equation}
	\mathbf{F}_{G_i}^{\alpha} = \exp(\alpha \widetilde{\mathbf{T}}_i),
\end{equation}
where (following the standard FRFT generator) 
\begin{equation}
	\widetilde{\mathbf{T}}_i 
	= -\mathrm{j}\,\frac{\pi}{2}\!\left(\pi(\mathbf{D}_{G_i}^2
	+ \mathbf{F}_{G_i}\mathbf{D}_{G_i}^2\mathbf{F}_{G_i}^{-1})-\tfrac{1}{2}\mathbf{I}\right).
\end{equation}
Its derivative is
\begin{equation}
	\frac{d}{d\alpha}\mathbf{F}_{G_i}^{\alpha}
	= \widetilde{\mathbf{T}}_i\mathbf{F}_{G_i}^{\alpha}.
\end{equation}
By the properties of the Kronecker product, it follows that
\begin{equation}
	\frac{\partial\mathbf{F}_{2D}^{(\alpha_1,\alpha_2)}}{\partial\alpha_1}
	= \mathbf{F}_{G_2}^{\alpha_2} \otimes \dot{\mathbf{F}}_{G_1}^{\alpha_1},
	\quad
	\frac{\partial\mathbf{F}_{2D}^{(\alpha_1,\alpha_2)}}{\partial\alpha_2}
	= \dot{\mathbf{F}}_{G_2}^{\alpha_2} \otimes \mathbf{F}_{G_1}^{\alpha_1}.
\end{equation}
The gradients of the loss with respect to $\alpha_1$ and $\alpha_2$ are then obtained via the chain rule.


\bibliographystyle{IEEEtran}
\bibliography{reference}

\newpage

\vfill

\end{document}